\documentclass[twocolumn,english,aps,prx,superscriptaddress,floats,nobibnotes]{revtex4-1}
\usepackage[latin9]{inputenc}
\usepackage{color}
\usepackage{babel}
\usepackage{bm}
\usepackage{bbm}
\usepackage{amsmath}
\usepackage{amssymb}
\usepackage{graphicx}
\PassOptionsToPackage{version=3}{mhchem}
\usepackage{mhchem}
\usepackage{braket}
\usepackage{appendix}
\usepackage{comment}
\usepackage[unicode=true,
 bookmarks=false,
 breaklinks=false,pdfborder={0 0 1},backref=false,colorlinks=false]
 {hyperref}
 \usepackage[normalem]{ulem}
\hypersetup{
 colorlinks,linkcolor=blue,citecolor=blue,urlcolor=blue}

\makeatletter

\usepackage{babel}
\PassOptionsToPackage{version=3}{mhchem}

\usepackage{newfloat}
\usepackage{latexsym}
\usepackage{dcolumn}
\usepackage{float}

\usepackage{bm}
\usepackage{babel}

\makeatother

\begin{document}

\title{Carrier transport theory for twisted bilayer graphene in the metallic regime}

\author{Girish Sharma}
\thanks{These two authors contributed equally to this work}
\affiliation{Centre for Advanced 2D Materials, National
University of Singapore, 6 Science Drive 2, 117546, Singapore}
\affiliation{School of Basic Sciences, Indian Institute of Technology Mandi, Mandi-175005, India}

\author{Indra Yudhistira}
\thanks{These two authors contributed equally to this work}
\affiliation{Centre for Advanced 2D Materials, National
University of Singapore, 6 Science Drive 2, 117546, Singapore}
\affiliation{Department of Physics, National
University of Singapore, 2 Science Drive 3, 117551, Singapore}

\author{Nilotpal Chakraborty}
\affiliation{Centre for Advanced 2D Materials, National
University of Singapore, 6 Science Drive 2, 117546, Singapore}
\affiliation{Yale-NUS College, 16 College Avenue West, 138527, Singapore}
\affiliation{Rudolf Peierls Centre for Theoretical Physics, Clarendon Laboratory, Parks Road, Oxford OX1 3PU, UK}

\author{Derek Y. H. Ho}
\affiliation{Centre for Advanced 2D Materials, National
University of Singapore, 6 Science Drive 2, 117546, Singapore}
\affiliation{Yale-NUS College, 16 College Avenue West, 138527, Singapore}

\author{Michael S. Fuhrer}
\affiliation{ARC Centre of Excellence in Future Low Energy Electronic Technologies, Monash University, Monash, Victoria  3800, Australia}
\affiliation{School of Physics and
Astronomy, Monash University, Monash, Victoria  3800, Australia}

\author{Giovanni Vignale}
\affiliation{Centre for Advanced 2D Materials, National
University of Singapore, 6 Science Drive 2, 117546, Singapore}
\affiliation{Yale-NUS College, 16 College Avenue West, 138527, Singapore}
\affiliation{Department of Physics and Astronomy, University of Missouri, Columbia, Missouri 65211, USA}

\author{Shaffique Adam}
\affiliation{Centre for Advanced 2D Materials, National
University of Singapore, 6 Science Drive 2, 117546, Singapore}
\affiliation{Department of Physics, National
University of Singapore, 2 Science Drive 3, 117551, Singapore}
\affiliation{Yale-NUS College, 16 College Avenue West, 138527, Singapore}

\date{\today}
\begin{abstract}
Understanding the normal-metal state transport in twisted bilayer graphene near magic angle is of fundamental importance as it provides insights into the mechanisms responsible for the observed strongly correlated insulating and superconducting phases. Here we provide a rigorous theory for phonon-dominated transport in twisted bilayer graphene describing its unusual signatures in the resistivity (including the variation with electron density, temperature, and twist angle) showing good quantitative agreement with recent experiments. We contrast this with the alternative Planckian dissipation mechanism that we show is incompatible with available experimental data.  An accurate treatment of the electron-phonon scattering requires us to go well beyond the usual treatment, including both interband and intraband processes,  considering the finite-temperature dynamical screening of the electron-phonon matrix element, and going beyond the linear Dirac dispersion. In addition to explaining the observations in currently available experimental data, we make concrete predictions that can be tested in ongoing experiments.      
\end{abstract}

\maketitle
The seminal observation of superconductivity and correlated insulating states in twisted bilayer graphene (tBG)~\cite{cao_unconventional_2018,cao_correlated_2018, yankowitz_2019_tuning} has generated tremendous excitement in the physics community \cite{gibney2019magic}.  At present, there is no consensus on the mechanism responsible for these observations.  It was anticipated almost 15 years ago \cite{Santos2007} that when two sheets of graphene are stacked on top of each other with a slight relative rotation, a large wavelength moir\'{e} superlattice potential emerges.  By reducing the twist angle in these moir\'{e} systems, the Bloch period can be increased by two orders of magnitude thereby bridging the lengthscales between naturally occurring lattices in materials and optical traps of cold atoms \cite{tarruell2012creating}.

The addition of a moir\'{e} potential significantly modifies the underlying electronic structure, including both a reduction in the Fermi velocity at low energy, and a reduction of the bandwidth of the lowest energy band.  Both these effects enhance the importance of electron-election interactions \cite{Tangscience2019}.  These properties can be understood as follows:  In the absence of any coupling between the layers, the original Dirac-like bands are just folded onto the smaller moir\'{e} Brillouin zone as determined by symmetry, but not modified; it is the interlayer coupling that causes level repulsion between the folded moir\'{e} bands. The moir\'{e} band closest to charge neutrality remains Dirac-like at low energy (the sublattice symmetry protecting the Dirac cones is not broken by the moir\'{e} potential) but with a reduced Fermi velocity; and the first moir\'{e} band as a whole gets squeezed by the level repulsion.  This reduced bandwidth is quantified by the separation in energy of the two van Hove singularities (VHS) that are found at the midpoint between the two original (relatively rotated in the Brillouin zone) Dirac cones, but pushed closer in energy by the level repulsion.  Numerical {\it ab initio} studies soon confirmed the predictions of this long-wave length continuum picture \cite{Trambly2010}; however, the experimental situation remained controversial for a while (see e.g. Ref.~\cite{Hicks2011}).  Since then, the continuum model has been largely confirmed experimentally (see e.g. Refs.\cite{LiNatphys2010,Wong2015}).

\begin{figure*}
\begin{center}
\includegraphics[height=!,width=16cm]{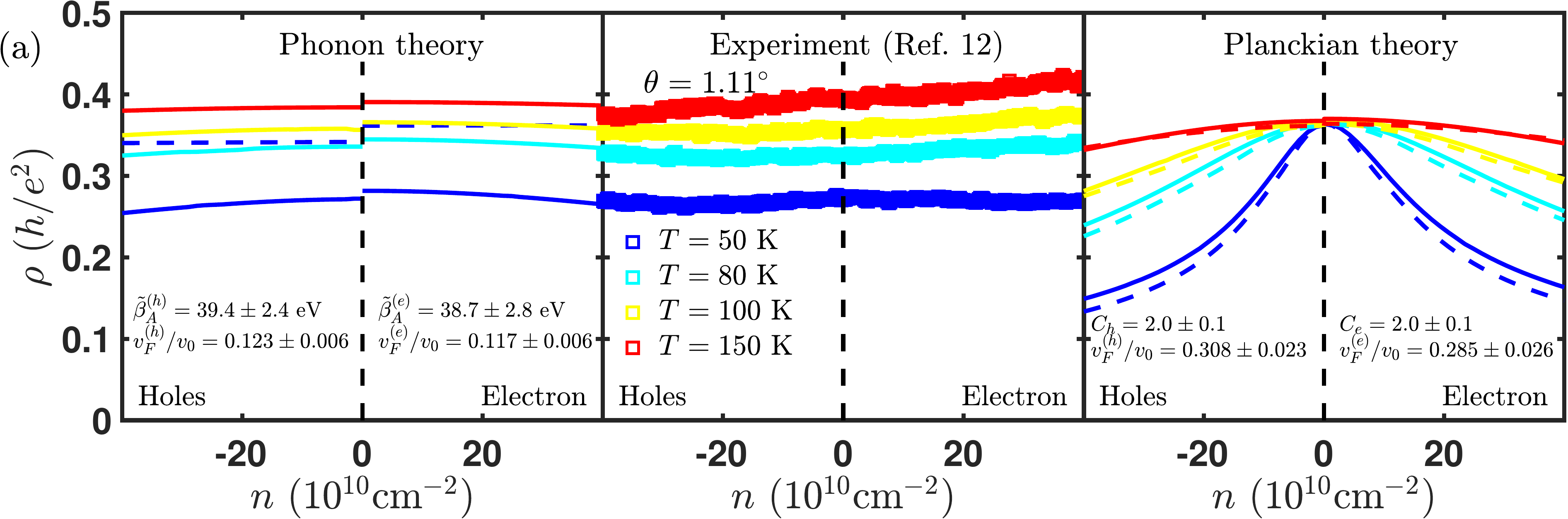}
\includegraphics[height=!,width=16cm]{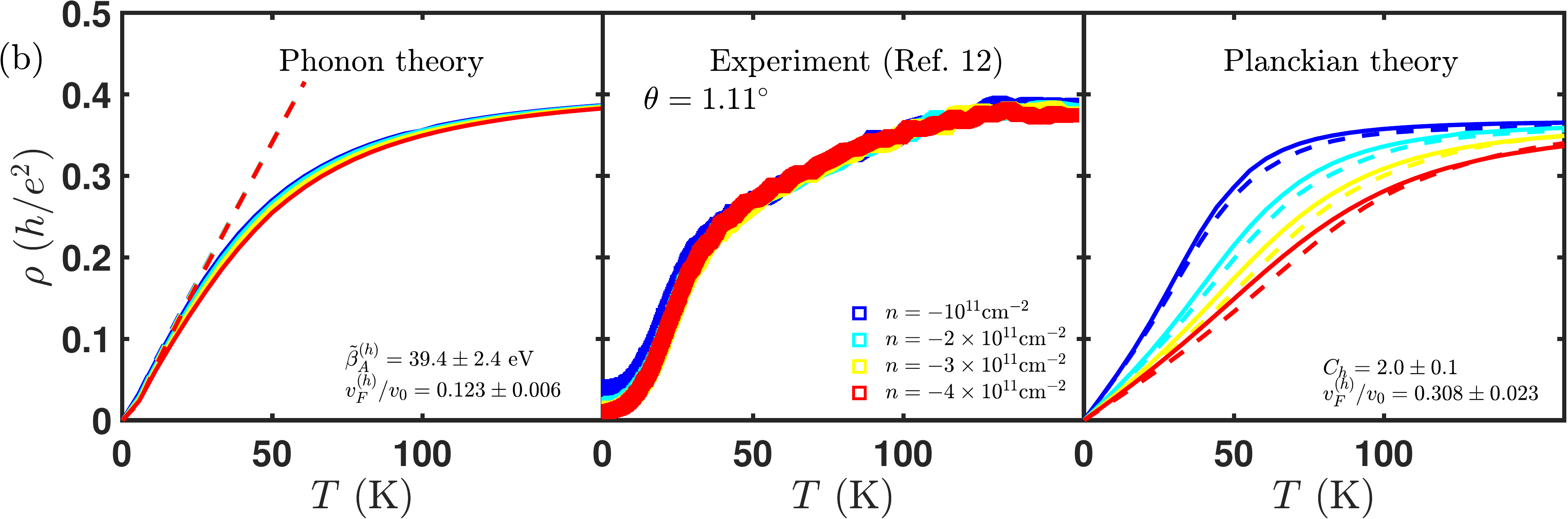}
\caption{The electron-phonon scattering theory (left panels) correctly captures the (a) carrier density and (b) temperature dependence of experimentally observed resistivity (middle panels), unlike the Planckian theory (right panel) that shows a stronger density dependence.  Experimental data is taken from Ref.~\cite{polshyn_2019_large} for $\theta = 1.11^\circ$ (comparison for devices with other twist angles is shown in Sec.~\ref{sec:expt}). Solid lines in the electron-phonon and Planckian theory are for a two-band effective model that includes the van Hove Singularity, while the dashed lines are for the linear Dirac model.  For electron-phonon scattering, the linear-in-$T$ resistivity at low temperature is captured by the Dirac model, while the saturation at higher temperature requires the van Hove singularity.  For the Planckian theory, the Dirac model and the two-band model are quantitatively similar and show much stronger density dependence compared to experiment.  In this case, the saturation at high-temperature is set not by the van Hove singularity, but by a universal value $\rho(T\rightarrow \infty) = C/8 \ln 2$, the coefficient $C \leq 1$ for Planckian dissipation).  For most experimental data, including those showed here, $C\geq 1$.  Taken together with the weak density dependence seen experimentally, this suggests that phonon scattering rather than Planckian dissipation is the dominant scattering mechanism at play in twisted bilayer graphene.}
\label{fig1}
\end{center}
\end{figure*}

Taking the continuum model to its logical conclusion, Bistritzer and MacDonald predicted \cite{bistritzer2011moire} that the Fermi velocity would vanish at a family of so-called ``magic angles''.  Their original work assumed that the lattices remained rigid.  More recent work including lattice relaxation effects \cite{jung2015origin,carr2019exact} suggests that only the first and largest magic angle ($\theta_M \sim 1.06^\circ$) is stable, and that the rigid lattice continuum approximation breaks down for smaller angles.  It should be emphasized that within the continuum model, strictly speaking, the bandwidth or $2\varepsilon_{\rm VHS}$ remains finite at the magic angle.  However, experimentally, at least in local spectroscopy measurements (e.g. Refs. \cite{jiang2019charge,kerelsky2019maximized,xie2019spectroscopic,choi2019electronic}), an 
alternate definition of magic angle is possible, i.e. when 
$\varepsilon_{\rm VHS}  =0$. These would occur at angles below the original magic angle and in the regime where lattice relaxation effects are dominant (and it is not clear, in this case, what the electronic structure would look like).  Given the observation of strongly correlated physics in other twisted 2D materials \cite{liu2019spin,tang2019wse2,wang2019magic,adak2020tunable}, it seems that the vanishing bandwidth is more germane than the vanishing Fermi velocity, although at present, the relation between the two has not been established.

In this work, we establish yet another special angle, $\theta_\mathrm{cr} \sim 1.15^\circ$, the angle at which the Fermi velocity equals the phonon velocity.  We show that at this angle, the phonon contribution to the resistivity strictly vanishes, and the experimentally measured resistivity would increase by several orders of magnitude for small deviations in angle on either side of $\theta_\mathrm{cr}$.  It has become normative in this quickly evolving field to attribute factor of $\sim~5$ changes in the resisitivity as evidence for superconductivity, and our work suggests more caution.  By construction, $\theta_\mathrm{cr}$ is larger than the original magic angle, and therefore its effects should be robust to lattice relaxation effects.  We demonstrate that the Fermi velocity of the linear bands and the van Hove singularities at the edges of the moir\'{e} Brillouin zone have distinct effects on the resistivity, and these could therefore be used in transport experiments to disentangle the importance of each in the correlated regime.

The present work is not about the observed superconductivity or correlated insulators.  As we explain here, there is a geometric enhancement of the electron-phonon coupling in such moir\'{e} systems \cite{choi_strong_2018, lian_twisted_2018} that would favour a phonon mechanism for superconductivity; however, in a separate paper \cite{sharma_collective_2019} we show that plasmons are also strongly enhanced and that superconductivity can arise from a purely electronic mechanism.  Similarly, at present it is unclear if the correlated insulator is a Mott insulator (see e.g. Refs.\cite{xu_topological_2018,roy_unconventional_2018,po_origin_2018,koshino_maximally_2018,kang_symmetry_2018,guo_pairing_2018,isobe_unconventional_2018,xie_on_2018,laksono2018singlet}) or a Wigner crystal \cite{padhi_doped_2018}. It is also unclear if the non-interacting bands are stable to the long-range Coulomb interaction~\cite{rademaker_charge-transfer_2018,guinea2018electrostatic} although experiments suggest they are. Rather, this work is about the carrier transport theory in the metallic regime (including at the van Hove singularity at higher carrier densities).  We find that the role of phonons in tBG is perhaps as interesting as that of electrons: the same moir\'{e} potential that gives rise to the flat electronic bands, also results in enhancement of the electron-phonon coupling. Soon after the first experiments~\cite{cao_unconventional_2018,cao_correlated_2018, yankowitz_2019_tuning}, we predicted~\cite{yudhistira2019gauge} that charged impurities would always dominate the resistivity at the lowest carrier densities and temperatures, but that gauge phonons would dominate for most of the experimental window. This crossover is also present in monolayer graphene, but occurs at a temperature of $\sim 500$~K, while for tBG the crossover happens at $\sim 5$~K.  As we show in Appendix~\ref{sec:expt}, available experimental data largely confirm our earlier predictions.  

By now there have been two experimental transport studies focusing on the metallic regime.  The first is from the MIT group \cite{cao_strange_2019} and the second is a UCSB-Columbia collaboration \cite{polshyn_2019_large}.  While the two experiments are largely consistent with each other, they arrive at very different conclusions on the dominant scattering mechanisms at play. Ref.~\cite{cao_strange_2019} argues for a Planckian mechanism to explain their data, which implies a scattering rate $\hbar \tau^{-1} = C k_{B} T$, where $C\lesssim 1$~\cite{bruin2013similarity}.  Here $C=1$ is the Planckian bound set by holography and believed to relevant for strange metals~\cite{hartnoll2018holographic}. They argue that the linear-in-temperature behaviour persisting well below the Bloch-Gruneisen temperature and the saturation of resistivity at higher temperature are both inconsistent with the conventional theory of phonon transport.  We show here that both of these features are actually essential features of phonon-limited transport in tBG.  While a microscopic theory showing Planckian dissipation has not yet been developed for tBG (it has for other systems, see e.g.~Ref.~\cite{tan2019realization,patel2019}), we can assume a Planckian mechanism to make predictions for the transport.  We show here that a Planckian mechanism also gives a saturation in resistivity at high temperature and linear-in-temperature behaviour at low temperature consistent with experimental observations.  However, a detailed analysis ultimately shows that the dominant scattering mechanism is not Planckian for several reasons including: (a) the Planckian theory also predicts a strong carrier density dependence (absent in the experiment); (b) the experiment and the phonon mechanism both show the resistivity saturation at high temperature is set by the VHS energy, while for the Planckian theory this saturation is intrinsic (i.e independent of bandstructure); (c) the twist angle dependence of Fermi velocity as extracted from experiment for phonon-limited scattering is consistent with the continuum theory~\cite{bistritzer2011moire,carr2019exact, jung2015origin}, while it is orders-of-magnitude off for the Planckian theory; and most significantly, (d) the extracted value of the scattering time from the experiment using the Planckian theory contradicts the assumptions of the Planckian theory.  Our work shows that the phonon interpretation of Ref.~\cite{polshyn_2019_large} is consistent with the theory we develop here.    

In Fig.~\ref{fig1} we compare data from Ref.~\cite{polshyn_2019_large} (middle panels) with both a phonon-limited theory (left-panel) and a Planckian theory (right panel).  Similar to the experimental data, the phonon-mediated theory has weak density dependence.   By contrast, the resistivity of the Planckian theory has strong density dependence (not seen in the experiment) that results from the density of states dependence of the Drude weight, which unlike electron-phonon, remains uncompensated by the scattering time.  We note that both the phonon-limited theory and the Planckian theory are linear-in-$T$ at low temperature, and saturate at high temperature (qualitatively similar to what is seen experimentally).  However, the origin of the saturation is very different.  For phonon scattering, the saturation is set by the electronic bandwidth $2\varepsilon_\mathrm{VHS}$, while for Planckian dissipation it is mostly independent of $\varepsilon_\mathrm{VHS}$ and set by Planckian strength $C$, which is expected to be somewhat universal and $C<1$. This illustrates that both the phonon-limited theory and the Planckian theory provide robust predictions that can be tested against experiment.  Details of how we fit the experimental data to electron-phonon and Planckian theory are provided in the supplemental material. 

Since phonons appear to dominate the transport properties, it is imperative that the theory is done correctly.  In this paper, we investigate the problem of normal state electronic transport in tBG focusing on the role of electron-phonon collision, which we find is the most important scattering mechanism in the relevant temperature and density regimes.  An accurate treatment of the electron-phonon scattering requires us to go well beyond the usual treatment, where by including both interband and intraband processes, we show, for example, that the interband process allows for a linear-in-$T$ behavior well below the Bloch-Gruneisen temperature, and the transition between the two is accompanied by several orders-of-magnitude decrease in the resistivity at a critical angle $\theta_\mathrm{cr}$, distinct from the magic angle $\theta_M$.  By considering the finite-temperature dynamical screening of the electron-phonon matrix element, we show, for example, that only the antisymmetric gauge phonon mode survives at low twist angle; and by going beyond the linear Dirac dispersion, we show that the van Hove singularity causes a saturation in resistivity as a function of temperature. In addition to explaining the observations in currently available experimental data, our theory makes concrete predictions that can be tested in ongoing experiments. 

\begin{figure}[h!]
    \centering
    \includegraphics[scale=0.21]{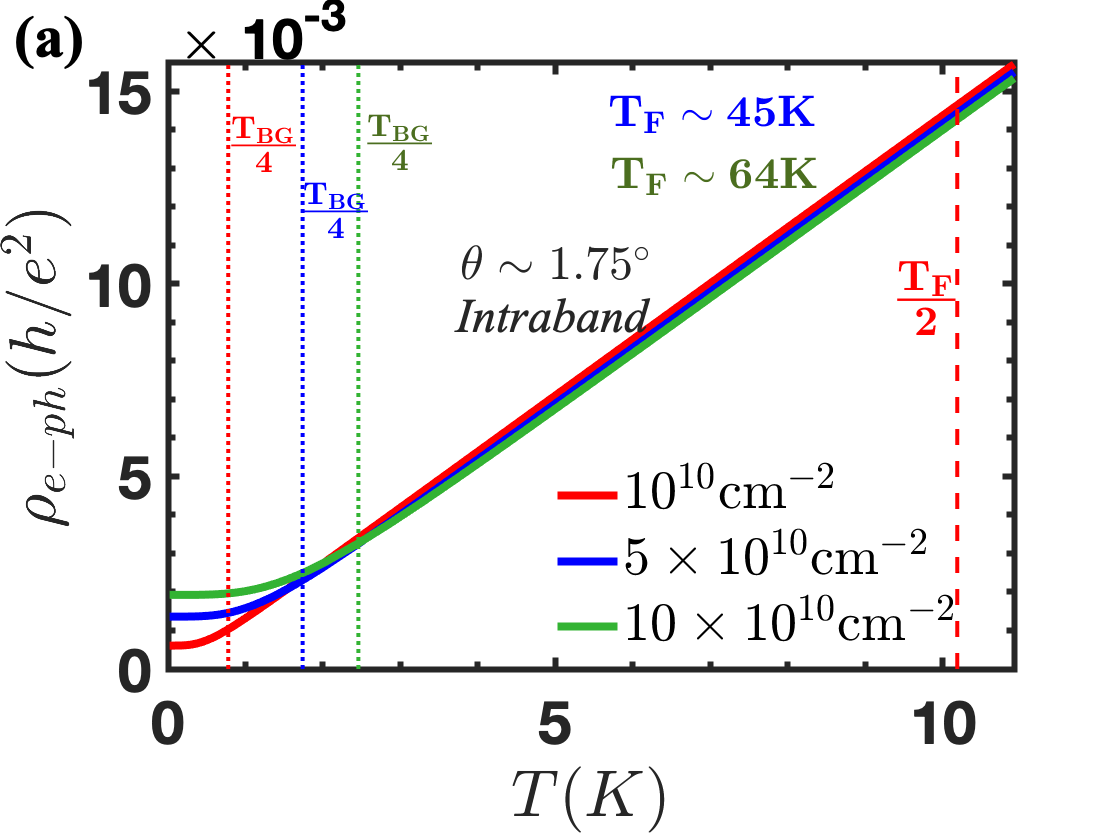}
     \includegraphics[scale=0.21]{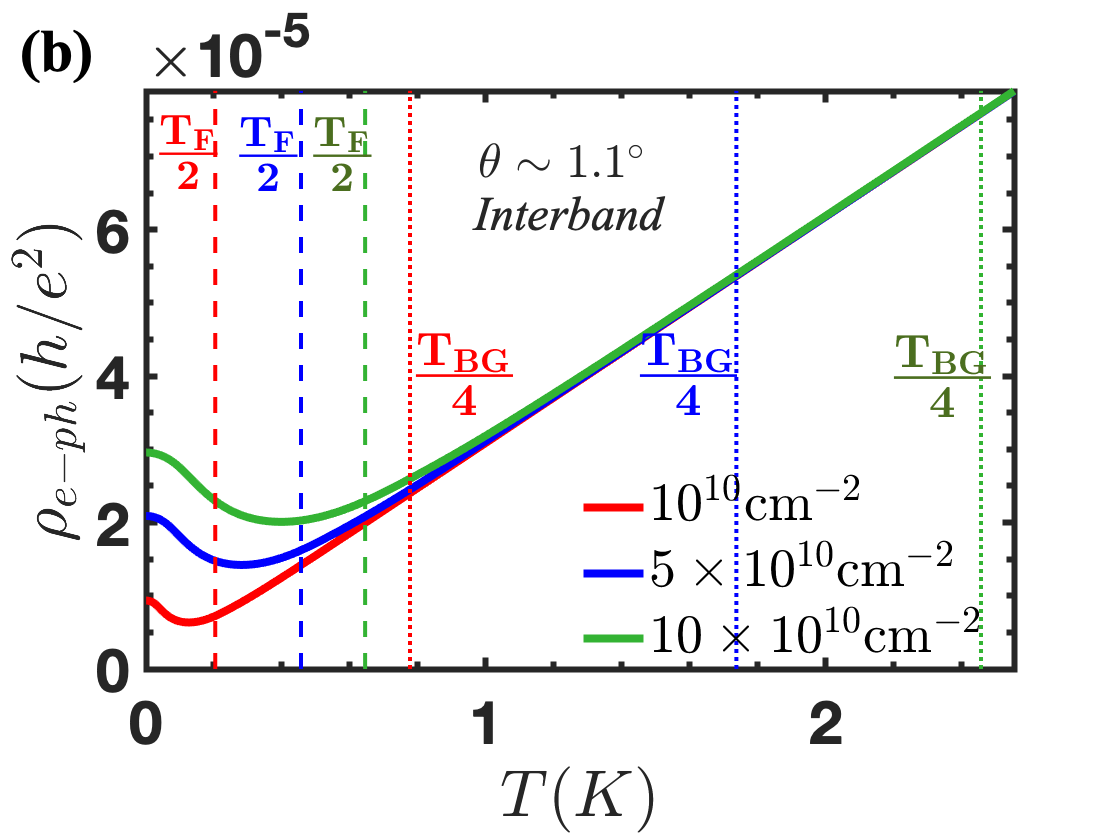}
    \caption{Electron phonon resistivity for tBG within the Dirac model for (a) intraband scattering, and (b) interband scattering. Dotted and dashed lines indicate $T_\mathrm{BG}/4$ and $T_F/2$ respectively. Interband resistivity shows a transition to linear-in-$T$ at $\sim T_F/2$ compared to intraband resistivity which shows a transition around $\sim T_\mathrm{BG}/4$. In the interband regime, the striking persistence of linear-in-$T$ behavior for $T\ll T_\mathrm{BG}$ is observed.}
    \label{Fig_rho_vs_T}
\end{figure}

To our knowledge, the Boltzmann transport theory for acoustic phonon scattering in monolayer graphene was first developed by Hwang and Das Sarma~\cite{Hwangacoustic2008}.  In this work, we adopt the same formalism with several extensions appropriate for twisted bilayer graphene.  First, we consider both interband and intraband processes.  As we show in Appendix \ref{sec:interintra}, close to the magic angle, only interband scattering is operational which was not considered in Ref.~\cite{Hwangacoustic2008}.  Second, while the linear Dirac Hamiltonian was an appropriate model for monolayer graphene, the reduced energies in tBG requires us to use an effective two-band Hamiltonian first proposed by Ref.~\cite{CNmodel} that captures the physics near the van Hove singularity.  Finally, we do the full finite-temperature and finite frequency RPA screening of the electron-phonon matrix elements (which is necessary due to the diverging density of states close at magic angle).  This demonstrates that it is the off-diagonal (or so-called gauge phonon contributions~\cite{suzuura2002phonons, sohier_phonon-limited_2014, yudhistira2019gauge}) of the acoustic phonon matrix element that dominate the transport properties.  How to screen the electron-phonon matrix element in two-dimensions has long remained controversial.  The issue is that prior to the present work, calculating the full dynamical polarazibility at finite temperature has been challenging.  Since the phonon propagator couples at a particular frequency, without dynamical screening it is unclear how the electrons screen the deformation potential.  This led to speculation in the theoretical literature as to whether the deformation potential should be screened or left unscreened.  For example, in Ref. \cite{Okuyama1989}, Okuyuma and Tokuda argue that experimental data for GaAs 2DEGs is better fit using the unscreeend theory, while in a later work, Kawamura and Das Sarma argue that once correctly done, static screening gives excellent agreement with experimental data \cite{Kawamura1992}.  Even for monolayer graphene, Ref.~\cite{sohier_phonon-limited_2014} argue that the deformation potential is completely screened, while Ref.~\cite{efetov2010controlling} argue for no screening.  In this work, we demonstrate conclusively that as anticipated by Kawamura and Das Sarma, the static screening approximation is closer to the correct dynamically screened result than the commonly used unscreened approximation.  Details of the screening are provided in Appendix~\ref{sec:dynamical}, and the geometric enhancement present for gauge phonons, but not scalar phonons is discussed in Appendix~\ref{sec:gauge}. 
\begin{figure}
\includegraphics[scale=0.4]{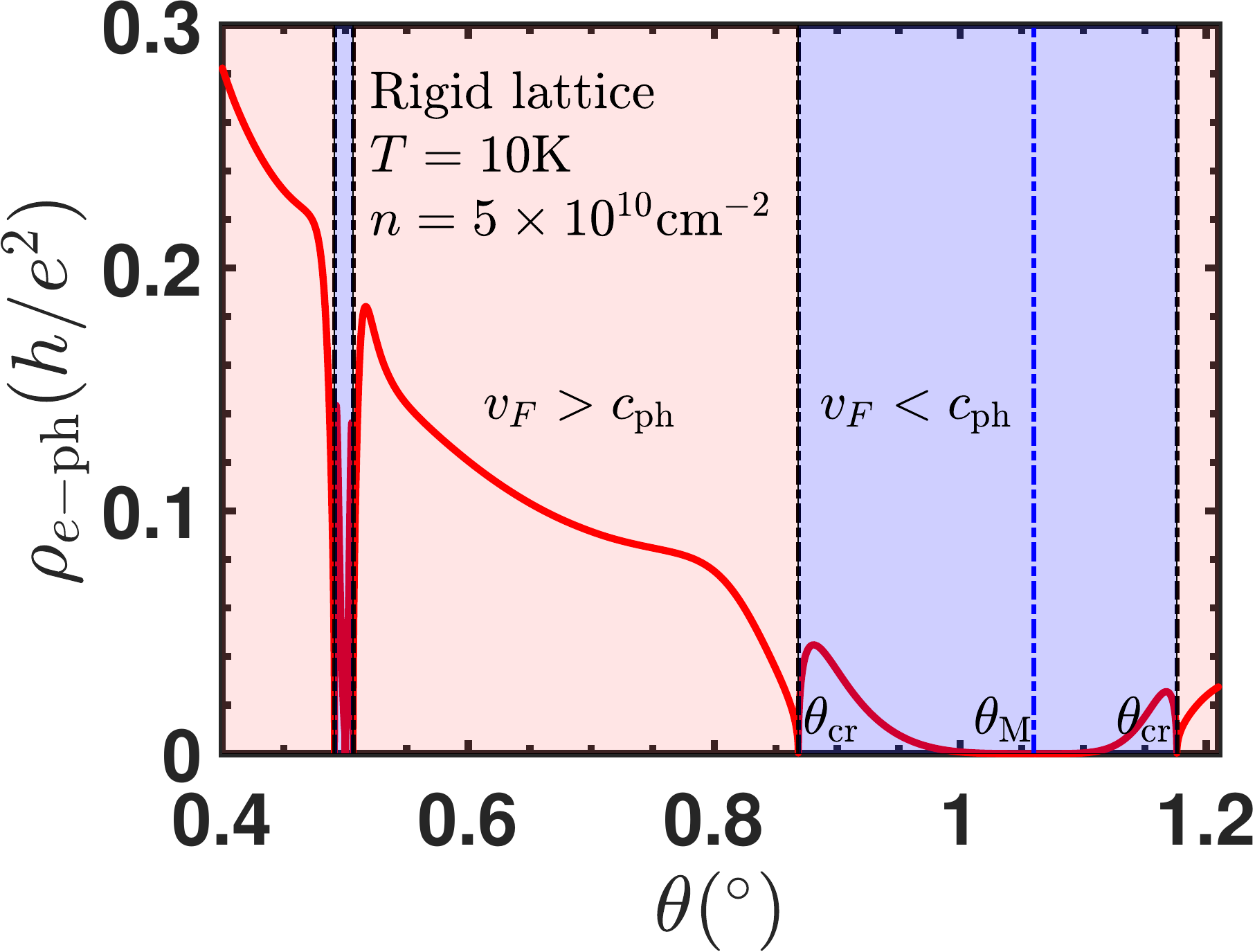}
\includegraphics[scale = 0.4]{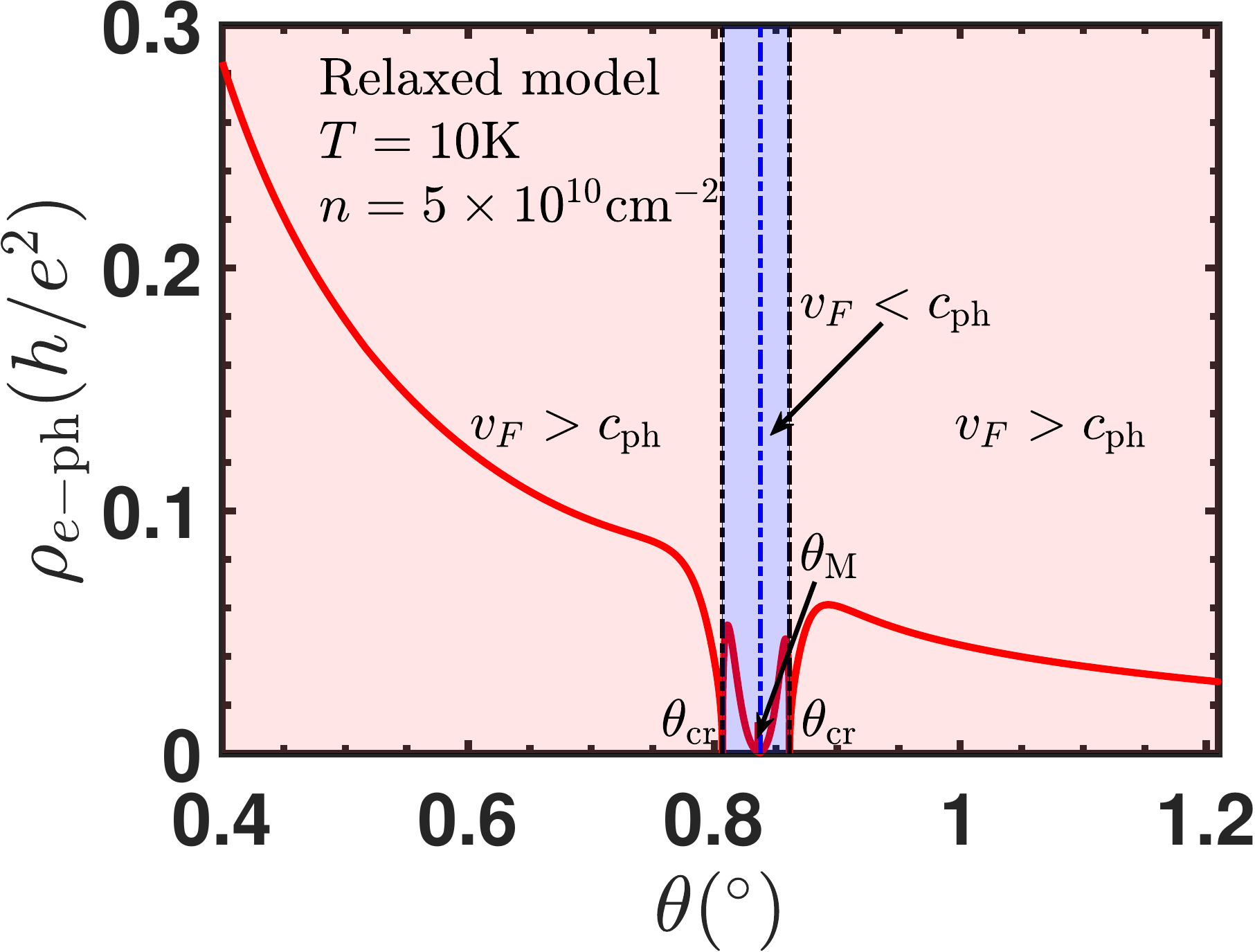}
\begin{center}
\end{center}
\caption{Close to the magic angle, the electron-phonon resistivity in tBG is very sensitive to the twist angle exhibiting a variation of several orders of magnitude when $v_{\rm F} = c_{\rm ph}$ and unrelated to Mott insulation or superconductivity.  The number of these sharp dips in resistivity and the angles at which they occur provide information about lattice relaxation~\cite{jung2015origin}.  (a) The rigid lattice model of Bistritzer and MacDonald~\cite{bistritzer2011moire} predicts that the resistivity will have multiple dips with decreasing twist angle (corresponding to three dips per magic angle).  (b) The relaxation model of Ref.~\cite{carr2019exact} gives only three sharp dips close to a single value of the magic angle.
The dashed black lines indicates the position of the critical angles $\theta_\mathrm{cr}$ when $v_F = c_\mathrm{ph}$. The dashed blue line indicates the position of the magic angle. The shaded red (blue) regions indicate regions when $v_F>c_\mathrm{ph}$ ($v_F<c_\mathrm{ph}$).
} \label{Fig:rhovstheta}
\end{figure}

{\it Interplay between Interband and Intraband scattering --}  There are two qualitatively distinct regimes depending on whether the phonon velocity $c_\mathrm{ph}$ is greater than or smaller than the Fermi velocity $v_F$. The crossover from $v_F>c_\mathrm{ph}$ to $v_F<c_\mathrm{ph}$ is expected because the renormalized $v_F$ vanishes at magic angles $\theta_M$.  For the largest magic angle, this crossover occurs at $\theta=\theta_\mathrm{cr} \sim 1.15^\circ$, and separates the regimes of interband ($v_F<c_\mathrm{ph}$) and intraband ($v_F>c_\mathrm{ph}$) scattering. Within the Dirac regime, the theory for intraband scattering is now well established~\cite{Hwangacoustic2008,efetov2010controlling,sohier_phonon-limited_2014, wu2018phonon}.  The resistivity shows a Bloch-Gr\"uneisen behaviour similar to metals and is given by $\rho_\mathrm{e-ph}=[16\zeta(\theta)^{2}k_{F}/(e^{2}\mu_s c_\mathrm{ph}v_F^{2})]F(T_{\mathrm{BG}}/T)$, where $T_\mathrm{BG}$ is the Bloch-Gr\"uneisen temperature ($T_\mathrm{BG} = 2\hbar c_\mathrm{ph} k_F$), $\mu_s$ is the graphene mass density, and  $F(x)=\int_{0}^{1}dy[xy^{4}\sqrt{1-y^{2}}e^{xy}]/\left(e^{xy}-1\right)^{2}$ (This form of the integral first appeared in Ref.~\cite{efetov2010controlling}).  For $T\gg T_{\rm BG}$, the quantization of the lattice phonon modes is irrelevant and the scattering is expected to be proportional to the amplitude of lattice vibrations, and is linear in T.

The intraband phonon scattering rate (dominant close to magic angle) shares some similarities with the interband scattering: it is density independent and $T-$linear at high-$T$.  Moreover, it vanishes when $v_F\rightarrow 0$ as the scattering phase space tends to zero.   However, qualitatively the interband and intraband scattering are quite different (see Appendix~\ref{sec:interintra}).  For example, while the intraband scattering rate within the Dirac model shows a monotonic increase with energy, the interband scattering rate is non-monotonic highlighting the suppression of interband scattering for energies larger than $k_B T$.  Most important, the temperature scale for the scattering rate to be $T-$linear is not set by the Bloch-Gr\"uneisen temperature $T_\mathrm{BG}$, but rather by the Fermi temperature $T_F$ (which is the maximum phonon energy at the Fermi surface allowed by kinematic constraints).  We note that $T_F$ and $T_\mathrm{BG}$ are defined in such a way so that at $\theta_\mathrm{cr}$, $T_F = T_\mathrm{BG}/2$.  Therefore, close to magic angle when $v_F<c_{ph}$, we have $T_F<T_\mathrm{BG}/2$ and the electron-phonon scattering becomes $T-$linear for temperatures well below $T_\mathrm{BG}$.  For $T\gg T_{\rm F}$, we find 
 \begin{equation}
\rho_\mathrm{inter}^{e-\mathrm{ph}}=\frac{h}{e^2}\frac{2\tilde{\beta}_{A}^{2}v_{F}^{2}k_{B}T}{\hbar^{2}\mu_{s}c_{\mathrm{ph}}^{6}},
 \end{equation}
where $\tilde{\beta}_A$ is the twist-angle dependent enhanced gauge field coupling constant.  Close to magic angle, we expect  $\rho_\mathrm{inter}^{e-\mathrm{ph}}(T\gg T_F)\propto v_F^4$ (where the additional $v_F^2$ comes from $\tilde{\beta}_A^2$, see Appendix~\ref{sec:gauge}), and vanishes at the magic angle due to the lack of scattering phase space.

Fig.~\ref{Fig_rho_vs_T}b shows the interband resistivity for a chosen $\theta<\theta_\mathrm{cr}$, comparing the scales of $T_F$ and $T_\mathrm{BG}$. The linearity in $T$ is observed to persist down to very low temperatures even when $T\ll T_\mathrm{BG}$, which is very different from the known theory of electron-phonon scattering in a typical Fermi liquid. The empirical observation of $T-$linear resistivity well below $T_\mathrm{BG}$ has been attributed to strange metallicity of non-Fermi liquids~\cite{cao_strange_2019}, however we find that there is nothing mysterious about this feature, it is merely the qualitative change in the nature of electron-phonon scattering when $v_F<c_\mathrm{ph}$ i.e. the lower of the two energy scales switches from lattice vibrational energy to electronic energy when $v_F$ crosses below $c_\mathrm{ph}$.

In Fig.~\ref{Fig:rhovstheta} we plot the electron-phonon resistivity for both the rigid continuum model~\cite{bistritzer2011moire} of tBG as well as including the lattice relaxation effects~\cite{carr2019exact}. We first note that whenever $v_F=c_\mathrm{ph}$, there are sharp dips in the resistivity profile, which can span a few orders of magnitude. Secondly at each magic angle there is another large dip.  For the rigid lattice model, there is a broad window where $v_F<c_\mathrm{ph}$ around the magic angle $\theta_M \sim 1.06^\circ$, and the family of magic angles implies multiple crossings of $v_F=c_\mathrm{ph}$, and hence multiple dips in the resistivity as the twist angle is lowered. With inclusion of lattice relaxation effects there is only a single stable magic angle, and there are therefore only three dips in the resistivity.  The window of the interband scattering regime ($v_F<c_\mathrm{ph}$) is also narrower.  We emphasize that close to magic angle, the resistivity is highly sensitive to twist angle exhibiting a variation of several orders of magnitude and one must be careful to experimentally distinguish this from Mott insulation or superconductivity.

{\it Beyond the Dirac model --}  While the Dirac model captures the physics of tBG at low density, in order to make accurate predictions for larger density and temperature, we need to extend the model to capture the VHS.  We use an effective two band Hamiltonian \cite{CNmodel}
\begin{equation}
    H(\textbf{k}) = - \frac{\hbar v_F}{|\Delta \mathbf{K}|} \begin{pmatrix}
    0 & k^{*2} - (\Delta \mathbf{K}^*/2)^2\\
    k^{2} - (\Delta \mathbf{K}/2)^2 & 0
    \end{pmatrix},
    \label{CN}
\end{equation}
 where $v_F$ is the Fermi velocity of twisted bilayer graphene, $k = k_x + ik_y$, $\Delta \mathbf{K}$ is wave vector separation between the two Dirac points which are located at $\mathbf{K}$ and $\mathbf{K}_\theta$, which magnitude is given by $k_\theta\equiv|\Delta \mathbf{K}|=2k_D\sin(\theta/2)$, with $k_D$ being the wave vector separation between the two Dirac points in monolayer graphene and $\theta$ is twist angle. The two band model is valid when the bandwidth is much larger than $\hbar v_F |\Delta \textbf{K}|$ which is a good approximation for small twist angles. The eigenenergies of this Hamiltonian are given by $\varepsilon_{\mathbf{k},\lambda}=\lambda(1/4)\hbar v_F\sqrt{k_\theta^2+8(k_x^2-k_y^2)+16[(k_x^2+k_y^2)/k_\theta]^2}$, and are anisotropic.  Within an isotropic approximation, we calculate the electron-phonon scattering rate for both intraband ($v_F > c_\mathrm{ph}$) and interband ($v_F < c_\mathrm{ph}$) with only intravalley electron-phonon scattering (See Appendix~\ref{sec:BT} for details).  We find
 
 \begin{figure*}
\includegraphics[scale=0.4]{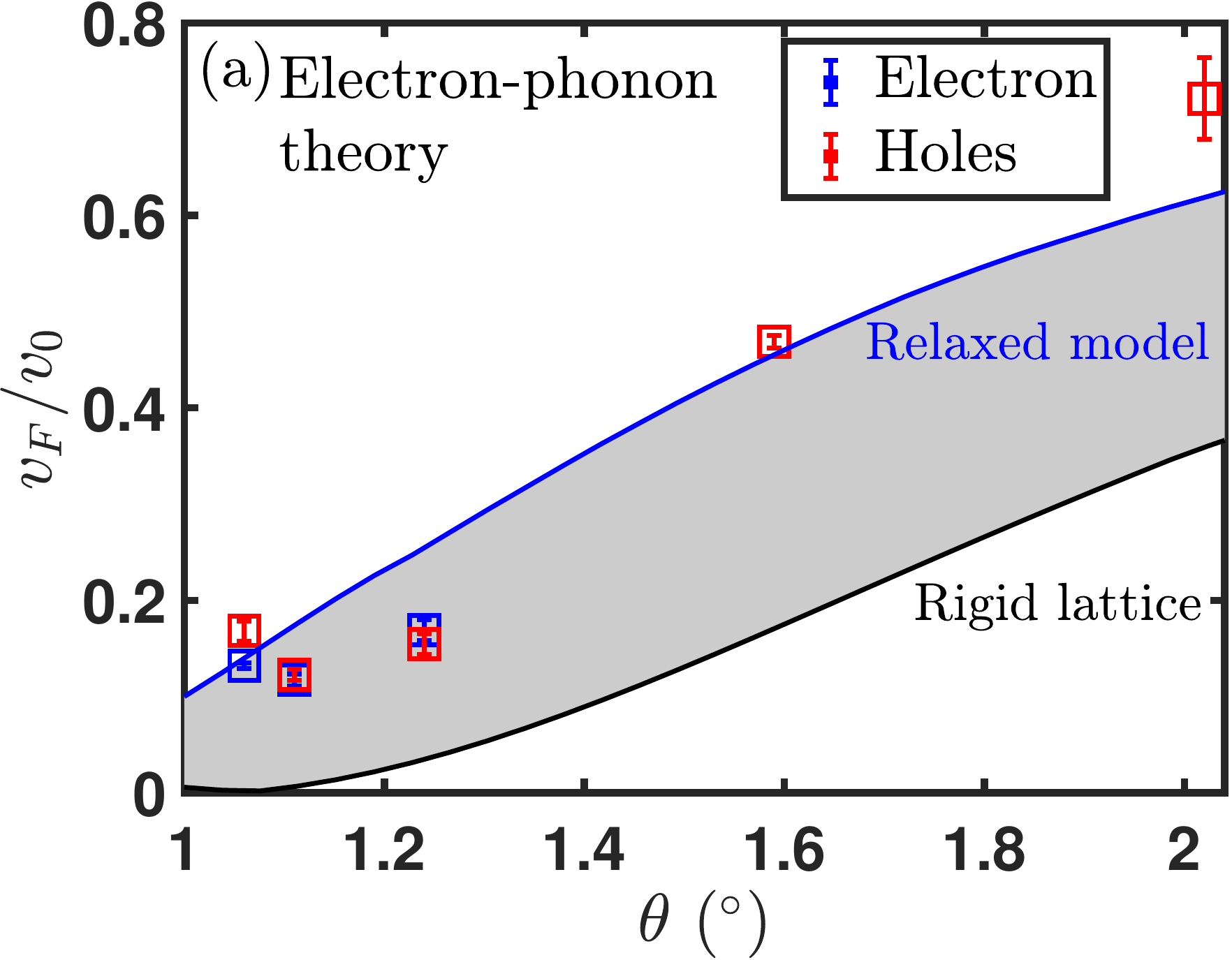}
\includegraphics[scale=0.4]{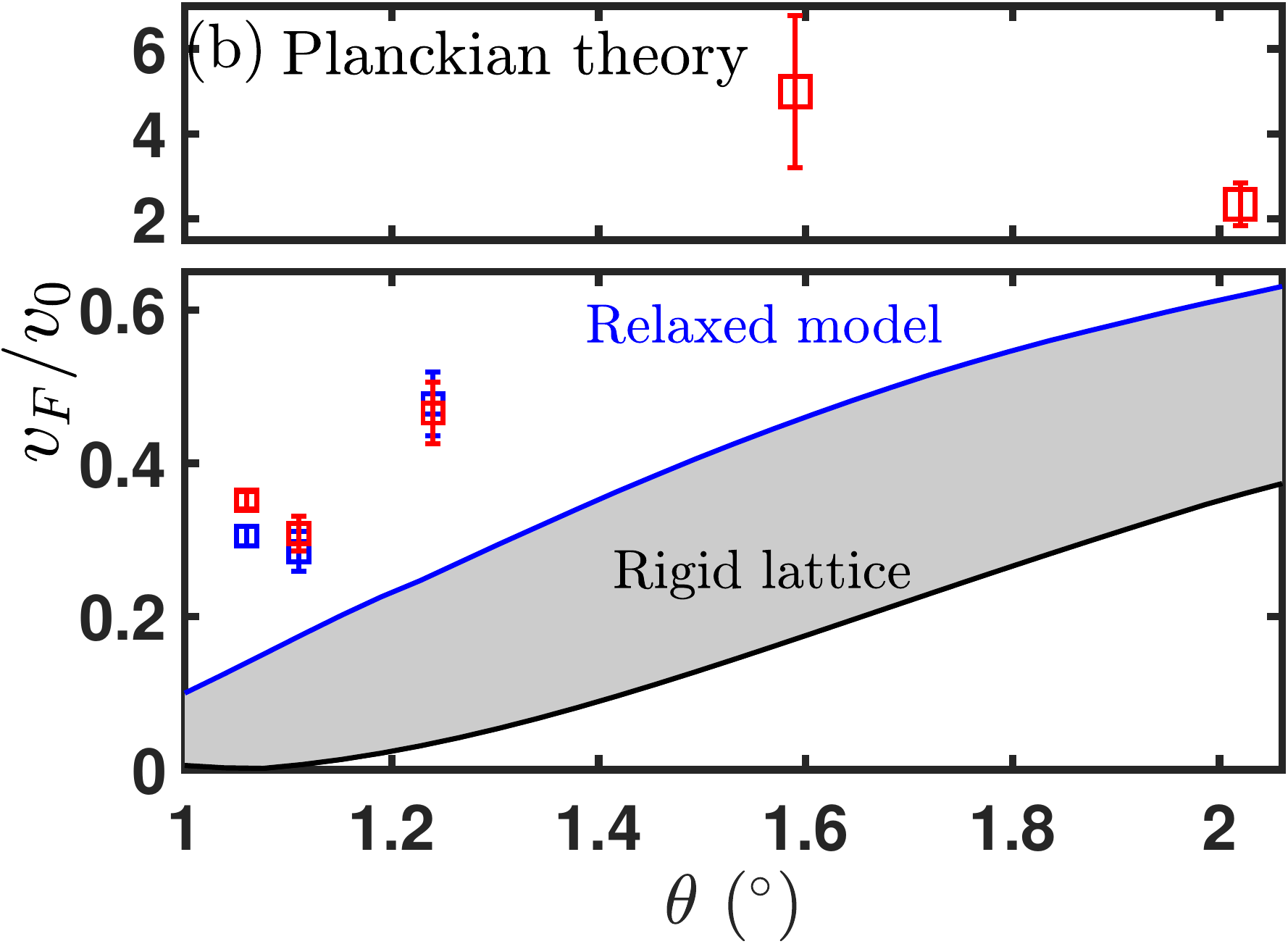}\\
\includegraphics[scale=0.4]{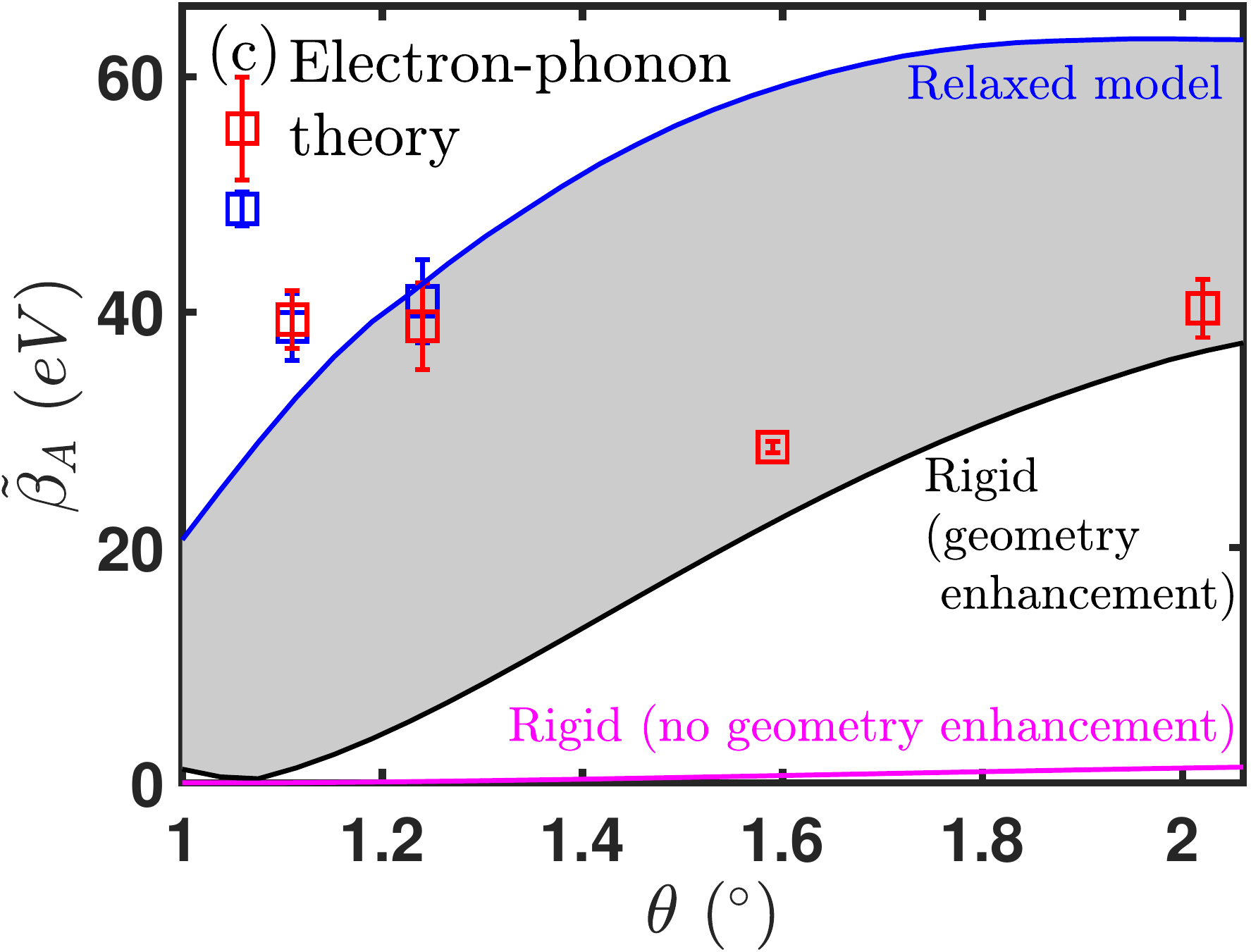}
\includegraphics[scale=0.4]{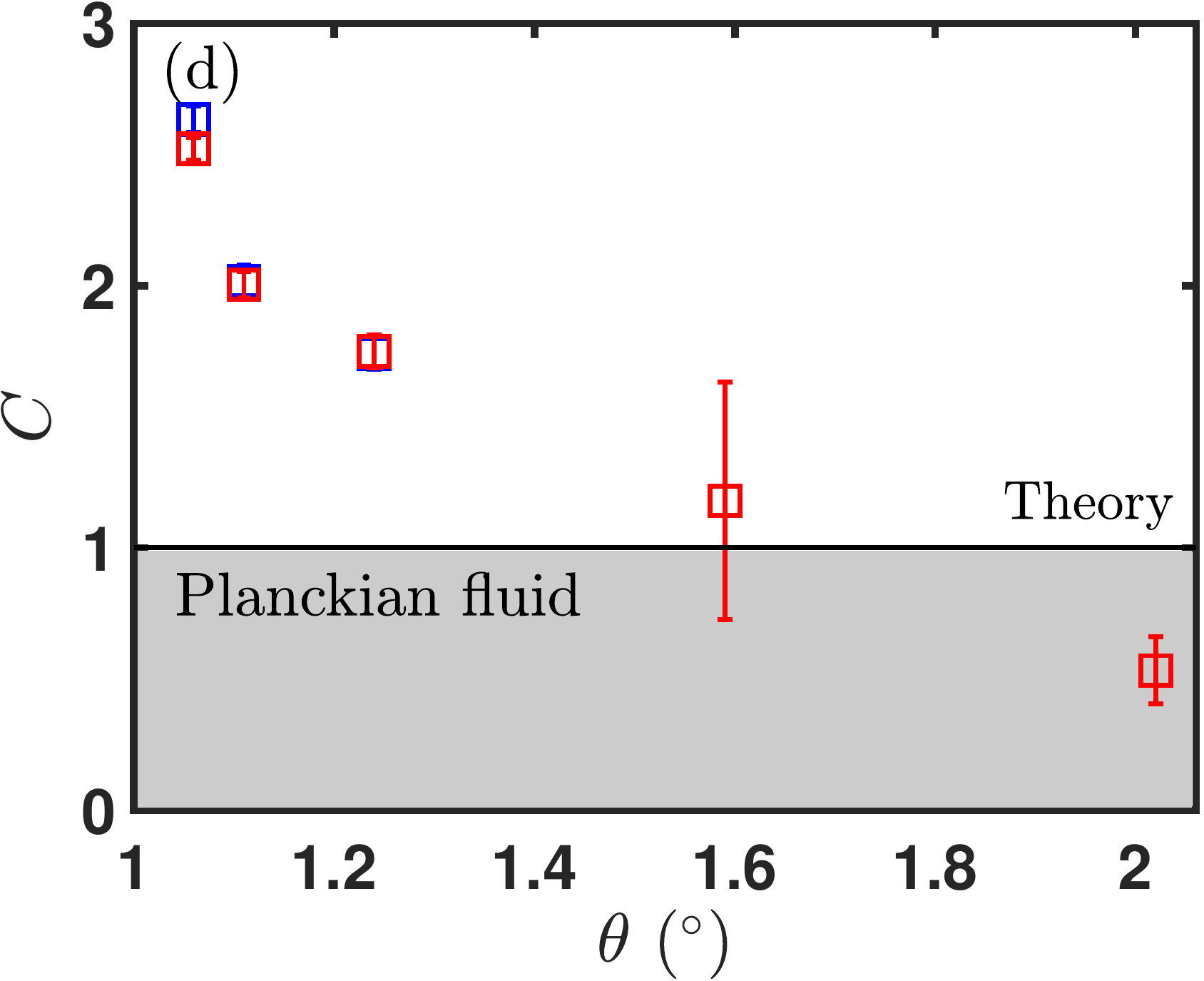}
\caption{(Color online) The parameters obtained from fit to Ref.~\cite{polshyn_2019_large} agree much better with the expectations from the electron-phonon theory than the Planckian model. (a) Fermi velocity as a function of twist angle obtained by fitting the experimental resistivity to the electron-phonon theory (b) obtained by fitting to the Planckian theory, (c) Effective electron-phonon coupling constant $\tilde{\beta}_A$ and (d) Planckian strength $C$. A violation of the Planckian bound ($C>1$) for small twist angles rules out the Planckian model as the dominant transport mechanism in tBG. \label{fig:parameter}}
\end{figure*}

 \begin{widetext}
\begin{align}
\label{eq:scattrateephCN}
\frac{1}{\tau_{\mathrm{intra}}^{e-\mathrm{ph}}(r)} &=\sum_{\substack{\xi=\pm1\\
\nu=\mathrm{LA,TA}}}\frac{\tilde{\beta}_{A}}{\pi\mu_{s}\hbar^{2}v_{F}c_{\nu}}\int_{0}^{rk_{\theta}/\left[2\left(1-\xi z_{\nu}\right)\right]}dqK_{\xi,\nu}\frac{q^{3}}{rk_{\theta}}\frac{16\sqrt{1-s_{\xi,\nu}^{2}}}{rk_{\theta}+4\xi z_{\nu}q}\left(\frac{1-\xi}{2}+\frac{1}{e^{\beta\hbar cq}-1}+\frac{\xi}{e^{\beta\left[(1/4)\hbar v_{F}k_{\theta}r+\xi\hbar c_{\nu}q-\mu\right]}+1}\right)\\
\frac{1}{\tau_{\mathrm{inter}}^{e-\mathrm{ph}}(r)} & =\sum_{\nu=\mathrm{TA},\mathrm{LA}}\frac{\tilde{\beta}_{A}}{\pi\mu_{s}\hbar^{2}v_{F}c_{\nu}}\int_{rk_{\theta}/\left[2\left(z_{\nu}+1\right)\right]}^{rk_{\theta}/\left[2\left(z_{\nu}-1\right)\right]}dqK_{-,\nu}\frac{q^{3}}{rk_{\theta}}\frac{16\sqrt{1-s_{-,\nu}^{2}}}{4z_{\nu}q-rk_{\theta}}\left(\frac{1}{e^{\beta\hbar c_{\nu}q}-1}+\frac{1}{e^{\beta\left[-(1/4)\hbar v_{F}k_{\theta}r+\hbar c_{\nu}q-\mu\right]}+1}\right),
\end{align}
 \end{widetext}
 where $z_\nu=c_\nu/v_{F}$, $s_{\pm,\nu}=[q/(2k)](z_\nu^{2}-1)\pm z_\nu$,
and $K_{\xi,\nu}=\sqrt{(1+r)^{2}+8(1+r)(q/k_{\theta})s_{\xi,\nu}+16(q/k_{\theta})^{2}}$. Here, $\xi= \pm 1$ refers to phonon absorption and emission, respectively. The resistivity for this Hamiltonian is
\begin{align}
\label{eq:rhoephCN}
\frac{1}{\rho_{ij}}  =8e^{2}\sum_{\lambda=\pm1}\int_{0}^{\infty}dr\int_{-\pi}^{\pi}&d\phi\frac{\mathcal{J}(r,\phi)}{(2\pi)^{2}}v_{\mathbf{k},\lambda}^{(i)}v_{\mathbf{k},\lambda}^{(j)}\nonumber\\
&\times\tau^{e-\mathrm{ph}}(r)\left(-\frac{\partial f^{0}}{\partial\varepsilon}\right)
\end{align}
where $v_{\mathbf{k},\lambda}^{(j)}=(1/\hbar)(\partial\varepsilon_{\mathbf{k},\lambda}/\partial k_j)$ is band velocity in $j$ direction. 

We plot the results of our calculation of $\langle\rho\rangle=\sqrt{\rho_{xx}\rho_{yy}}$ for twist angle of $\theta=1.1^\circ$ in the left panel of Fig.~\ref{fig1}. The electron-phonon resistivity is linear-in-$T$ at low temperature but saturates at high temperature. The slope of resistivity with temperature in the low temperature regime is set by $v_F$ while the saturation of resistivity with temperature is set by the bandwidth. Within this effective Hamiltonian, the bandwidth $2\varepsilon_\mathrm{VHS}$ and $v_F$ are not independent,  and related by $2\varepsilon_\mathrm{VHS}=(1/2)\hbar v_F k_\theta$. While this relation between $\varepsilon_\mathrm{VHS}$ and $v_F$ is specific to our model, we believe that the main conclusions are generic i.e.~the low temperature linear-in-$T$ behaviour is set by the Fermi velocity, while the saturation is set by the VHS. The saturation of electron-phonon resistivity can be simplified deep in the intraband ($v_F\gg c_\mathrm{ph}$) and interband ($v_F\ll c_\mathrm{ph}$) regime as
\begin{align}
\rho_{\mathrm{intra}}^{e-\mathrm{ph}}(T\rightarrow\infty) & \propto\frac{\tilde{\beta}_{A}^{2}}{e^{2}\hbar\mu_{s}}\frac{1}{v_{F}^{2}c_\mathrm{ph}^{2}}\varepsilon_{\mathrm{VHS}}\\
\rho_{\mathrm{inter}}^{e-\mathrm{ph}}(T\rightarrow\infty) & \propto\frac{\tilde{\beta}_{A}^{2}}{e^{2}\hbar\mu_{s}}\frac{v_{F}^{2}}{c_\mathrm{ph}^{6}}\varepsilon_{\mathrm{VHS}}
\end{align}

{\it Resistivity from Planckian model -- }  We take a simple phenomenological model with $\hbar \tau_\mathrm{Pl}^{-1} = C ~k_B T$.  
The resistivity is obtained from the Boltzmann equation by using the relaxation time approximation, with $\tau_{Pl}$ as the relaxation time (see discussion in Ref.~\cite{bruin2013similarity}), and
calculating the appropriate thermal average with the density of states and Fermi velocity (see Eq.~\ref{eq:rho_formula}).     We can do this both for the Dirac model and the two-band Hamiltonian. Surprisingly, both the Dirac Hamiltonian and the two-band model give very similar results indicating that the van Hove singularity is not important for the Planckian theory.  This phenomenological model exhibits a linear-in-$T$ resistivity at low temperature that saturates at higher temperature (qualitatively similar to what is seen experimentally).  We find that the the slope of resistivity with temperature in the low temperature regime is set by both $v_F$ and $C$, while the saturation of resistivity with temperature is set only by $C$.  Within the Dirac approximation, the resistivity is 
\begin{align}
    \frac{1}{\rho_{\mathrm{Pl}}} = \frac{1}{C}\frac{4e^2}{h}\sum\limits_{\lambda=\pm 1} \ln \left[1+ \exp\left(\lambda\frac{\mu}{k_B T}\right)\right],
\end{align}
\noindent with low and high temperature asymptotes   
\begin{align}
\rho_\mathrm{Pl}=\begin{cases}
C\frac{h}{4e^{2}}\frac{T}{T_{F}}\left[1+\frac{\pi^{2}}{6}\left(\frac{T}{T_{F}}\right)^{2}\right] & ;T\ll T_{F}\\
C\frac{h}{e^{2}}\frac{1}{8\ln2}\left[1-\frac{1}{128\left(\ln2\right)^{3}}\left(\frac{T_{F}}{T}\right)^{4}\right] & ;T\gg T_{F}
\end{cases}
\end{align}

For the effective two band model, the Planckian resistivity is given by an expression similar to Eq.~\ref{eq:rhoephCN}.  Introducing dimensionless variables $\tilde{\mathcal{J}}=\mathcal{J}/k_\theta^2$, $\tilde{\mu} = \mu/k_B T$, and $\tilde{v}_{\mathbf{k},\lambda}^{(j)}=\partial\tilde{\varepsilon}_{\tilde{\mathbf{k}},\lambda}/\partial\tilde{k}_{j}$, where $\tilde{\varepsilon} _{\tilde{\mathbf{k}},\lambda}=\varepsilon_{\mathbf{k},\lambda} /\varepsilon_\mathrm{VHS}$ and $\tilde{\mathbf{k}}= \mathbf{k}/k_\theta$, it can be simplified to
 \begin{align}
\frac{1}{\rho_{\mathrm{Pl}}^{ij}} & =\frac{e^{2}}{h}\frac{1}{C}K_{j}\left(\frac{n}{n_{\mathrm{VHS}}},\frac{k_{B}T}{\varepsilon_{\mathrm{VHS}}}\right),
\end{align}
where the function $K_{j}$ is computed numerically (see Appendix~\ref{sec:BT}). In the right panel of Fig.~\ref{fig1}a and \ref{fig1}b, we show the Planckian resistivity as a function of density and temperature, respectively at twist angle of $1.11^\circ$.  The Planckian resistivity for both models saturate at $\rho_\mathrm{Pl}(T\rightarrow\infty) = C / 8\ln 2$, independent of the bandwidth $\varepsilon_\mathrm{VHS}$, which is in sharp contrast to electron-phonon scattering.   Figure.~\ref{fig:parameter} shows the results of fitting 8 data sets of varying twist angle, temperature and carrier density to both the phonon and Planckian models (full fits are shown in the Appendix).  We find that the phonon-scattering theory (but not the Planckian model) gives fit parameters for both the Fermi velocity and model parameters that are consistent with theoretical expectations.

The nature of electron-phonon scattering investigated here gives rise to several new features: Interband electron-phonon scattering, which should be kinematically forbidden in monolayer graphene, is shown to occur in tBG below a critical twist angle $\theta_\mathrm{cr}$, when $v_F<c_\mathrm{ph}$. The critical angle ($\theta_\mathrm{cr}$) is a sweet spot where both the interband and intraband scattering phase space (and thus resistivity) both drop to zero, giving rise to multiple dips in the resistivity unrelated to insulating or superconducting states that could be investigated in the future experiments. We derive explicit analytical expressions for the interband scattering rate and show its qualitative dissimilarity from the intraband scattering.  Importantly, we show that this explains the linear-in-$T$ resistivity well below the Bloch-Gruneisen temperature $T_\mathrm{BG}$, a previously unexplained experimental puzzle.

In this work we also provide additional theoretical verification of our earlier claim~\cite{yudhistira2019gauge} that the gauge phonon modes are enhanced by the moir\'{e} geometry and are not screened, while the scalar phonon modes have neither property.  We also provide experimental verification (see Fig.~\ref{Fig:rho_exp_Dirac_vsn} and \ref{Fig:rho_exp_Dirac_vsT} in Appendix ~\ref{sec:expt}) of our previous predictions that charged impurity scattering takes over as the dominant scattering mechanism at very low temperatures (below 20K) and low carrier densities (below $\sim 10^{11}~{\rm cm}^{-2}$).  Taken together with the present work on phonon scattering (that applies at high temperature and high carrier density), this now presents a complete theory for the carrier transport for twisted biayer graphene.   

\begin{acknowledgements}
We acknowledge the Singapore Ministry of Education AcRF Tier 2 Grant No. MOE2017-T2-2-140, the National
University of Singapore Young Investigator Award (Grant No. R-607-000-094-133), and use of the dedicated research computing resources at CA2DM. MSF and SA acknowledge support of the ARC through grants CE170100039 and DP200101345.  It is a pleasure to thank M.~M.~E.~Alezzi, C. Dean, E.~Laksono, P. Jarillo-Herrero H.~Mahalingam, N. Raghuvanshi, and M. Yankowitz for fruitful discussions. 
\end{acknowledgements}

\bibliographystyle{apsrev4-1}
\bibliography{biblio.bib}

\clearpage

\appendix
\begin{widetext}

\section{Interplay between Inter-band and Intra-band Scattering}
\label{sec:interintra}
The Dirac approximation of tBG consists of a degenerate Dirac cone with a renormalized Fermi velocity $v_F$ which is heavily suppressed near the magic angle $\theta_M$. Even though this approximation is valid only below the Van Hove singularity (VHS), it nevertheless allows us to analytically examine the interesting qualitative features in the electronic transport of tBG.
We show that the renormalization of the Fermi velocity has important implications on phonon assisted scattering mechanisms particularly in the regime when the Fermi velocity becomes close to the sound velocity, a situation which is rather unusual in typical metals.  
Since both $\theta_M$ and $\theta_\mathrm{cr}$ are very close, the magic angle physics becomes even more interesting with the interplay of electrons and phonons, which gives specific signatures in the electronic transport.  While $v_F$ can be controlled by twist angle,  $c_\mathrm{ph}$ is relatively insensitive to changes in the twist angle.

From the best estimates available for LA phonons in graphene, bilayer graphene, twisted bilayer graphene~\cite{cocemasov2013phonons} and graphite, we expect the effective phonon velocity $c_\mathrm{ph}$ defined by $2/c_\mathrm{ph}^2=1/c_\mathrm{LA}^2+1/c_\mathrm{TA}^2$ in the current scenario to lie in the range of $c_\mathrm{ph}\sim$ 20-30 km s$^{-1}$, and not heavily sensitive to changes in the twist angles (see Table~\ref{table:1}). It has also been shown that the tBG phonon spectrum is largely insensitive to the details of moir\'{e} superlattice even at small twist angles close to 1$^{\circ}$~\cite{choi_strong_2018}.   

\begin{table}[H]
\caption{Comparison of phonon sound velocity in various Carbon based materials.} 
\centering{}%
\begin{ruledtabular}
\begin{tabular}{cccc} 
System & Sound velocity (km/s)  \tabularnewline
\hline\tabularnewline
monolayer graphene & $\sim$ 20~\cite{efetov2010controlling,chen2008intrinsic,sohier_phonon-limited_2014}\tabularnewline
bilayer graphene & $\sim$ 20~\cite{cocemasov2013phonons} \tabularnewline
graphite & $\sim$ 21~\cite{ono1966theory}\tabularnewline
carbon nanotubes & $\sim$ 21~\cite{suzuura2002phonons}\tabularnewline
twisted bilayer graphene & $\sim$ 20~\cite{cocemasov2013phonons} \tabularnewline
\end{tabular}
\end{ruledtabular}
\label{table:1}
\end{table}

We now discuss the kinematics of electron-phonon scattering focusing on small values of the renormalized Fermi velocity. Specifically two types of processes are possible. An electron in band $\lambda$ with momentum $\mathbf{k}$ can scatter into an electron in band $\lambda'$ with momentum $\mathbf{k}\pm \mathbf{q}$ due to emission or absorption of a phonon. The case $\lambda=\lambda'$ ($\lambda \neq \lambda'$) correspond to intra (inter) band scattering respectively. Additionally we also have the requirement of energy conservation $\varepsilon_{\mathbf{k}\pm \mathbf{q},\lambda'} = \varepsilon_{\mathbf{k},\lambda} \pm \hbar c_\nu q$.  
We can visualize the conditions for intraband and interband phonon scattering process diagrammatically as shown in Fig.~\ref{Fig:phononscattering}. If $\mathbf{k}_\mathrm{in}$ is the initial wave vector, and $\mathbf{q}$ is the phonon wavevector, then $k_\mathrm{in}-(c_\mathrm{ph}/v_F)q$ denotes the locus of allowed final states as dictated by conservation of energy for intraband scattering, while $(c_\mathrm{ph}/v_F)q - k_F$ is the locus of final allowed states for interband scattering, indicated by a solid circle. The locus of the final states dictated by momentum conservation is indicated by the dotted circle should intersect with the solid circle for scattering to take place. Based on this observation we can conclude that an intraband phonon scattering process is kinematically forbidden if $c_\mathrm{ph}>v_F$, while interband scattering is forbidden if $c_\mathrm{ph}<v_F$. As one approaches magic angle, one necessarily crosses over from intraband to the intraband regime. 

The above kinematics has implications on the electron-phonon scattering rate. The electron-phonon scattering time for the Dirac model of tBG can be solved within the Boltzmann formalism (see Appendix~\ref{sec:BT} for calculation details). In the limit $v_F\rightarrow c_\mathrm{ph}$ (i.e.~at $\theta\rightarrow\theta_\mathrm{cr}$) we specifically evaluate from Eq.~\ref{Eq:elecphonon1} that $1/\tau^{e-\mathrm{ph}}_\mathrm{inter} (\varepsilon,T)\rightarrow 0$ as well as $1/\tau^{e-\mathrm{ph}}_\mathrm{intra} (\varepsilon,T)\rightarrow 0$, highlighting that the scattering phase space exactly vanishes at this critical point. 
The electron-phonon scattering time can be further simplified deep in the intraband regime ($v_F\gg c_\mathrm{ph}$) and large-$T$ ($k_B T\gg \hbar \omega_\mathbf{q}$) to be
\begin{align}
        \frac{1}{\tau^{e-\mathrm{ph}}_\mathrm{intra} (\varepsilon,T)} = \sum_{\nu=\mathrm{LA,TA}}\frac{1}{4} \left(\frac{|\varepsilon| \zeta(\theta)^2}{\hbar^2 \mu_s }\right) \left(\frac{1}{c_\nu^2v_F^2}\right) \left(\frac{k_B T}{\hbar}\right).
        \label{Eq:tauintraanalytic1}
\end{align}
The scattering rate is $T-$linear and density independent \textcolor{black}{at high-$T$}, consistent with earlier findings in the literature~\cite{kaasbjerg2012unraveling}. 

The interband regime, which is of interest especially in the vicinity of the magic angle remains unexplored so far {in the literature}.
The electron-phonon scattering rate can also be simplified deep in the interband regime ($v_F \ll c_\mathrm{ph}$) and at \textcolor{black}{high}-$T$  ($k_B T\gg \hbar \omega_\mathbf{q}$) to be (see Appendix~\ref{sec:BT})
\begin{align}
    \frac{1}{\tau^{e-\mathrm{ph}}_\mathrm{inter} (\varepsilon,T)} =\sum_{\nu=\mathrm{LA,TA}}  \left(\frac{|\varepsilon| \zeta(\theta)^2}{\hbar^2 \mu_s}\right) \left(\frac{v_F^2}{c_\nu^6}\right) \left(\frac{k_B T}{\hbar}\right).
    \label{Eq:tauinteranalytic1}
\end{align}
Fig.~\ref{Fig_Tau_vs_E} plots the interband and intraband scattering rates as a function of energy. The numerical results closely agree with the above analytical expressions at high-$T$. The usual intraband scattering is linear in temperature for $T>T_\mathrm{BG}/4$ and $E\ll k_BT$. Further, we find that the interband scattering is also linear in temperature, but for $T> T_F/2$. This explains how phonon scattering can give linear-in-$T$ behavior at temperatures well below $T_{\rm BG}$.

\begin{figure}
\begin{center}$
\begin{array}{cc}
\includegraphics[height=!,width=4cm]{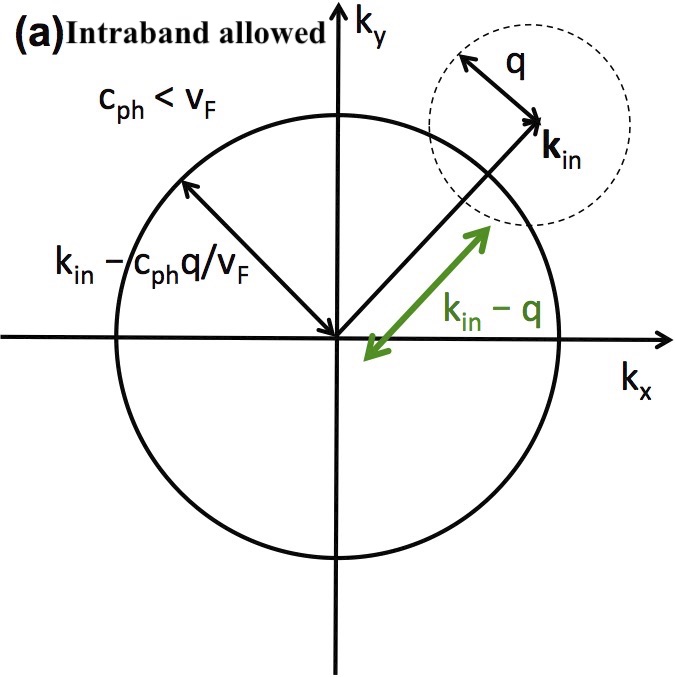} &
\includegraphics[height=!,width=4cm]{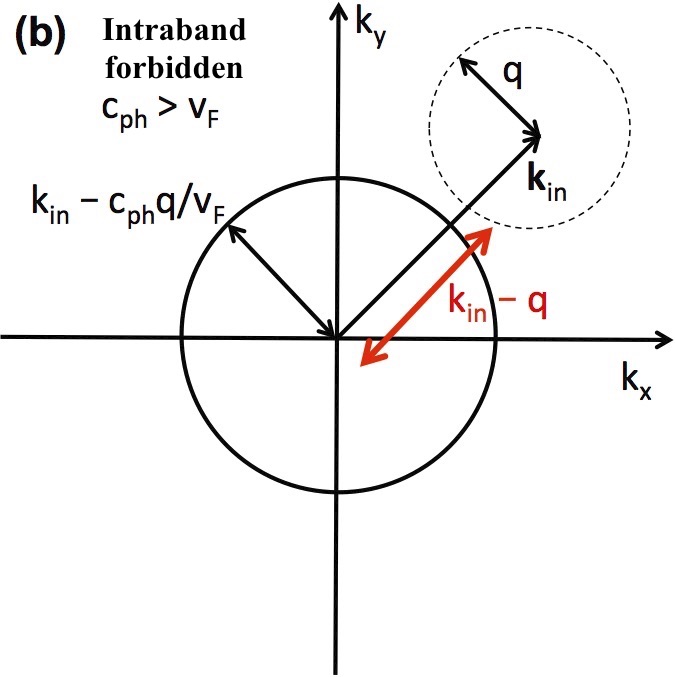} \\
\includegraphics[height=!,width=4cm]{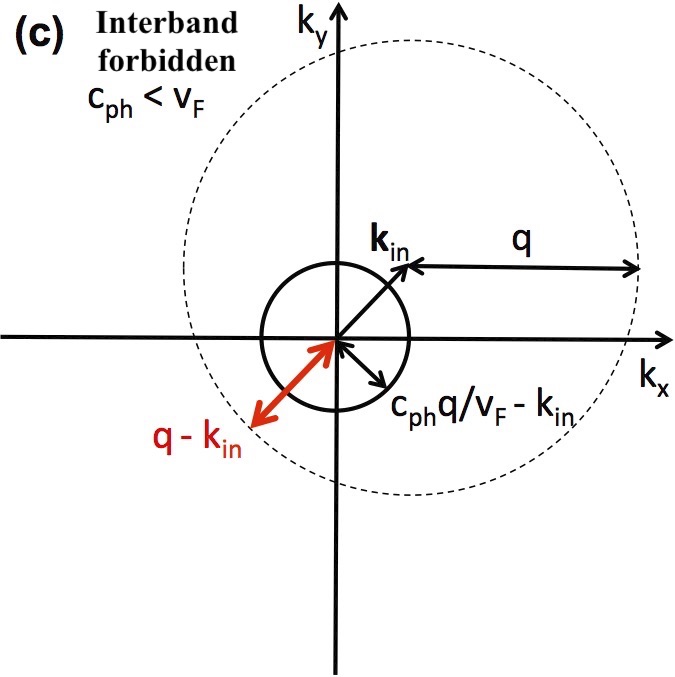} &
\includegraphics[height=!,width=4cm]{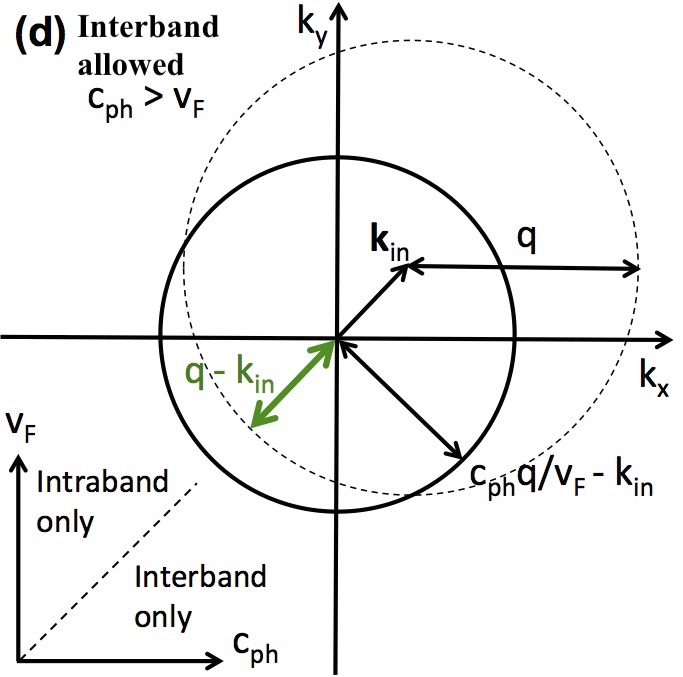} \\
\end{array}$
\end{center}
\caption{Kinematic constraints imply that only intraband phonon scattering is allowed for Fermi velocity larger than the phonon velocity ($v_F>c_\mathrm{ph}$) and only interband scattering is allowed for $v_{F}<c_\mathrm{ph}$.  
The solid (dashed) circle displays the locus of final electronic momenta allowed by conservation of energy (momentum).
Intraband electronic transitions are kinematically allowed when (a) $c_\mathrm{ph} < v_{F}$ and forbidden when (b) $c_\mathrm{ph} > v_{F}$. 
On the other hand, interband electronic transitions are kinematically forbidden when (c) $c_\mathrm{ph} < v_{F}$ and allowed when (d) $c_\mathrm{ph} > v_{F}$. In all panels, $k_{\mathrm{in}}$ refers to the initial electronic wave vector and $q$ to the phonon wave vector.
Inset: Depending on the relationship between $c_\mathrm{ph}$ and $v_{F}$, either intraband or interband transitions are allowed. Similar statements can be proven for phonon absorption using similar kinematic diagrams.  In twisted bilayer graphene, Fermi velocity decreases with twist angle, and we define the critical angle $\theta_\mathrm{cr}$ where $v_{F}=c_\mathrm{ph}$.  Here, the electron-phonon scattering length diverges resulting in several orders of magnitude decrease in the resistivity.  As discussed in the main text, we estimate $\theta_\mathrm{cr} \approx 1.15^\circ$ which is always larger than the magic angle defined here as when $v_{\rm F}$ vanishes.} \label{Fig:phononscattering}
\end{figure}

\begin{figure}
    \centering
    \includegraphics[scale = 0.2]{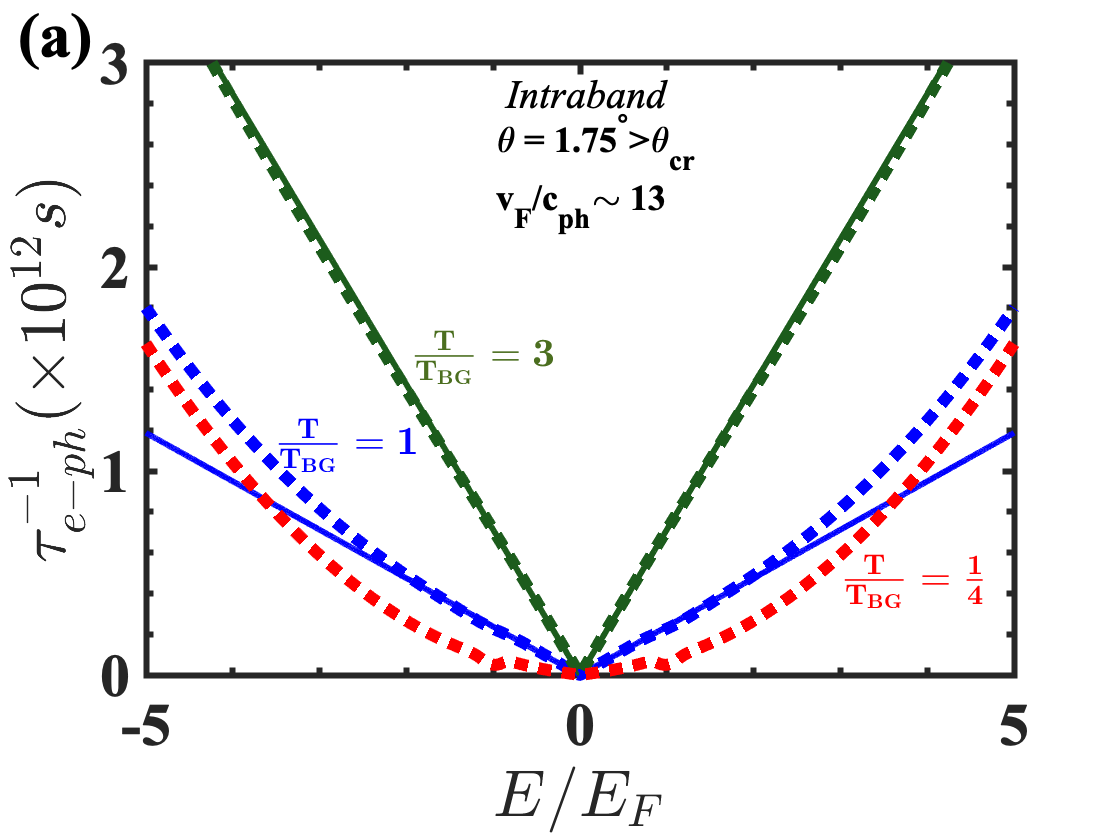}
    \includegraphics[scale=0.2]{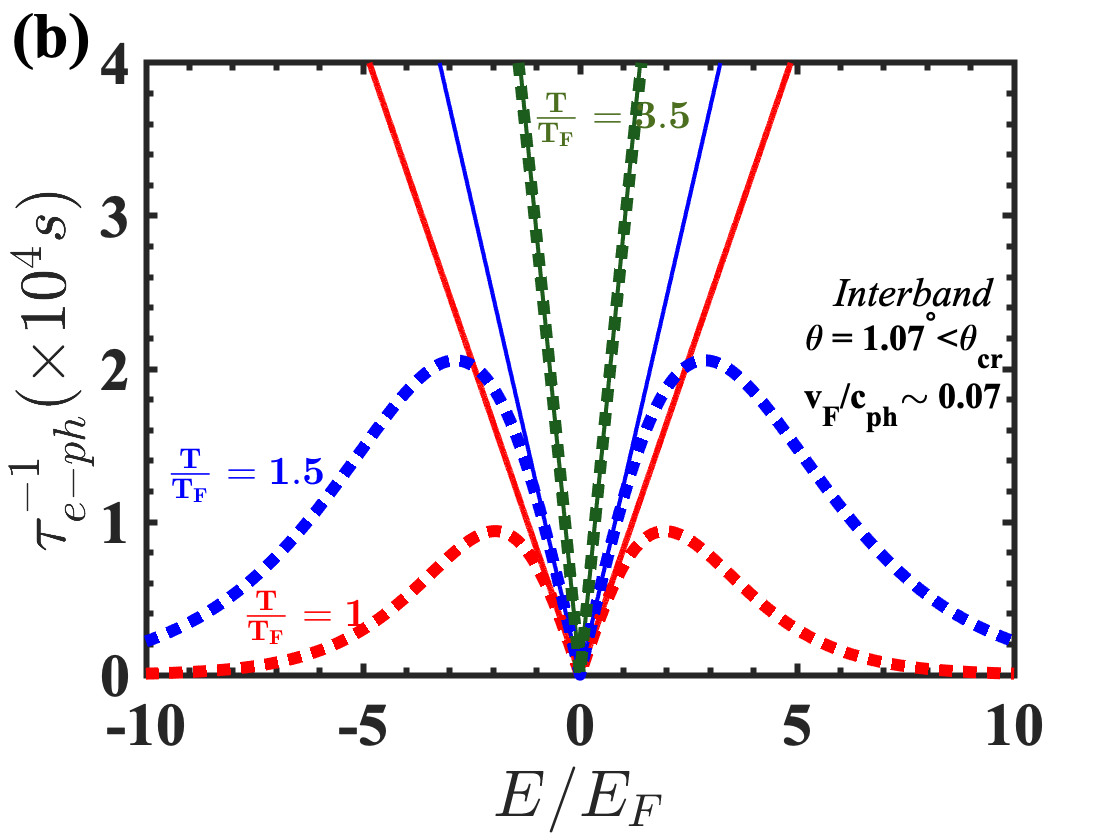}
    \caption{Interband and intraband phonon scattering are qualitatively different. Data points are a numerical solution of Eq.~\ref{Eq:elecphonon1}. (a) The usual intraband scattering is linear in temperature for $T>T_\mathrm{BG}/4$ and $E\ll k_BT$, (solid lines are the analytical expression, Eq.~\ref{Eq:tauintraanalytic1}).  (b) We find that the interband scattering is also linear in temperature, but for $T> T_F/2$.  This explains how phonon scattering can give linear-in-$T$ behavior at temperatures well below $T_{\rm BG}$.  Solid lines are the analytical result, Eq.~\ref{Eq:tauinteranalytic1}.  We note that while intraband scattering becomes stronger for $E\gg k_BT$, interband scattering becomes weaker in the same limit.}
    \label{Fig_Tau_vs_E}
\end{figure}

\section{Dynamical Screening of Electron-phonon coupling}
\label{sec:dynamical}

The role of screening is crucial in a phonon dominated carrier scattering theory, and as we shall see it is even more non-trivial in the case of tBG. Within a simple Thomas-Fermi screening model, we qualitatively note that the divergence of the dielectric constant $\epsilon^{-1} \sim v_F$ for vanishing Fermi velocities is exactly canceled by the the $v_F^{-1}$ dependence of resistivity for most scattering mechanisms including the scalar phonon modes~\cite{yudhistira2019gauge}. Gauge phonons on the other hand are unaffected by screening, which dramatically enhances its importance close to the magic angle (see Appendix.~\ref{sec:gauge} for more details on gauge and scalar phonon modes in tBG). Particularly, the crossover temperature for which gauge phonons dominate over charged impurities drops from room temperature in monolayer graphene to the order of few Kelvins in tBG close to the magic angle. This conclusion holds even when we consider static screening calculated within the random phase approximation (RPA) i.e. $\epsilon(\mathbf{q},\omega\rightarrow 0)$. The effect of frequency dependence of $\epsilon(\mathbf{q},\omega)$ on screening of the electron-phonon vertex has not been considered so far. This is because typically static screening is a good approximation in metals when screening due to impurities or phonons is considered. However, the frequency dependent dielectric function $\epsilon(\mathbf{q},\omega)$ is essential to describe properties like the dynamic screening, which becomes essential in tBG because electronic and phononic energy scales are quantitatively very similar. Recently we have discussed the role of $\epsilon(\mathbf{q},\omega)$ on superconductivity in tBG~\cite{sharma_collective_2019}.
Here we show the dynamic screening properties of scalar deformation potential phonon modes within the RPA. 

The basic building block of RPA screening is the polarizability bubble $\Pi_C(\mathbf{q},\omega)$, which can be written in the most general form as 
\begin{align}
  \Pi_C (\mathbf{q},i\omega_m) =\frac{1}{\beta} \sum_{\mathbf{k},i\omega_n} {\text{Tr}[G(\mathbf{k},i\omega_n) G(\mathbf{k} +\mathbf{q},i\omega_n +i\omega_m)] },
  \end{align}
where the summation $i\omega_n$ is over the imaginary frequency, $G(\mathbf{k},i\omega)$ is the Green's function, and the trace is over the sublattice degrees of freedom. Performing the trace and the Matsubara summation we obtain
  \begin{align}
      \Pi_{C}(\mathbf{q},i\omega)=\lim_{\eta\rightarrow0^{+}}\sum\limits _{\lambda,\lambda',\mathbf{k}}\frac{F_{\mathbf{k},\mathbf{k}+\mathbf{q}}^{\lambda\lambda'}}{2}\frac{f_{\mathbf{k},\lambda}^{0}-f_{\mathbf{k}+\mathbf{q},\lambda'}^{0}}{\hbar\omega+\varepsilon_{\mathbf{k},\lambda}-\varepsilon_{\mathbf{k}+\mathbf{q},\lambda'}+i\eta}     \label{Eq_Pi_q_omega}
  \end{align}
where $F^{\lambda\lambda'}_{\mathbf{k},\mathbf{k}+\mathbf{q}} = 1+\lambda\lambda' \cos\theta_{\mathbf{k},\mathbf{k}+\mathbf{q}}$ is the graphene chirality factor. The dielectric function $\epsilon(\mathbf{q},i\omega_m)$ is given by the RPA summation and is related to the basic pair bubble as $\epsilon(\mathbf{q},\omega) = 1-V_\mathbf{q}\Pi_C(\mathbf{q},i\omega)$, where $V_\mathbf{q} = 2\pi e^2/\kappa q$ is the Fourier transform of the Coulomb potential, $\kappa$ being the dielectric constant. We evaluate $\Pi_C$ and $\epsilon(\mathbf{q},\omega)$ for finite frequencies and temperatures, semi-analytically. To the best of our knowledge, this has not been done previously in the literature, at least in the context of phonons. 
The electron-phonon vertex is renormalized as $ g_\mathbf{q} \rightarrow {g_\mathbf{q}}/{\epsilon(\mathbf{q},\omega)}
$. 
In Fig.~\ref{Fig:epsilon_qomega1}a we plot the dynamic dielectric constant $|\epsilon(q,\omega)|$ for monolayer graphene and tBG at small twist angles, and discuss its implications on screening of the electron-phonon vertex $g_\mathbf{q}$.  The static screening case (long wavelength limit) corresponds to the limit of $\omega\rightarrow 0$, where $\epsilon(\mathbf{q})$ is always real and greater than unity.  Therefore the electron-phonon coupling is always screened compared to the bare value as expected. Dynamic screening, on the other hand, must be considered separately in the interband ($v_F<c_{ph}$) and intraband regimes ($v_F>c_{ph}$). 
In the intraband regime, the phonon spectrum always lies in the region when $|\epsilon(q,\omega)|>1$, and therefore the vertex $g_\mathbf{q}$ in this case will be always screened.
In the interband regime, the phonon spectrum intersects regions where $|\epsilon(q,\omega)|$ can be both greater or smaller than one. This is because the finite frequency dielectric function can become imaginary with its absolute value $|\epsilon(\mathbf{q},\omega)|$ becoming less than one for certain values of $\mathbf{q}$ and $\omega>v_F q$.
When $|\epsilon(q,\omega)|<1$, the vertex $g_\mathbf{q}$ is in fact anti-screened, in which case the electron-phonon coupling will be enhanced rather than suppressed.
For the interband regime, the effect on the scattering rates and resistivity need to be explicitly calculated (since these involve an integral over all possible wavevectors).  We find that the for intraband scattering, the dynamic screening and static screening give results that are qualitatively and quantitatively similar; however, for interband scattering, the full dynamical screening is necessary.  
Another feature which is evident from Fig.~\ref{Fig:epsilon_qomega1}b is that the phase space when $|\epsilon(q,\omega)|<1$ is reduced in tBG due to the suppressed Fermi velocity. This can be understood qualitatively by specifically searching for the zeros of the dielectric function ($|\epsilon(q,\omega)|=0$), which gives us the plasmon spectrum $\omega_{pl}$. The low-$q$ limit of the plasmon spectrum $\omega_{pl} = \sqrt{g e^2 E_F q/2\kappa}$. Clearly, $q/k_F = (2\kappa v_F/ge^2) (\omega_{pl}/E_F)^2$. Therefore if we fix $\omega/E_F$, the dimensionless wavevector $q/k_F$ decreases with decrease of $v_F$. This argument can be extended to the general case when $|\epsilon(q,\omega)|<1$, and the shrinking of blue region in Fig.~\ref{Fig:epsilon_qomega1} is expected. 
Fig.~\ref{fig:epscut} plots $|\epsilon(q,\omega)|$ for graphene and tBG (within the Dirac model) at different temperatures and frequencies. Again, the suppression of anti-screening and enhancement of screening effects in tBG is highlighted.

\begin{figure}[h]
  \centering
  \includegraphics[height=!, width=8.5cm]{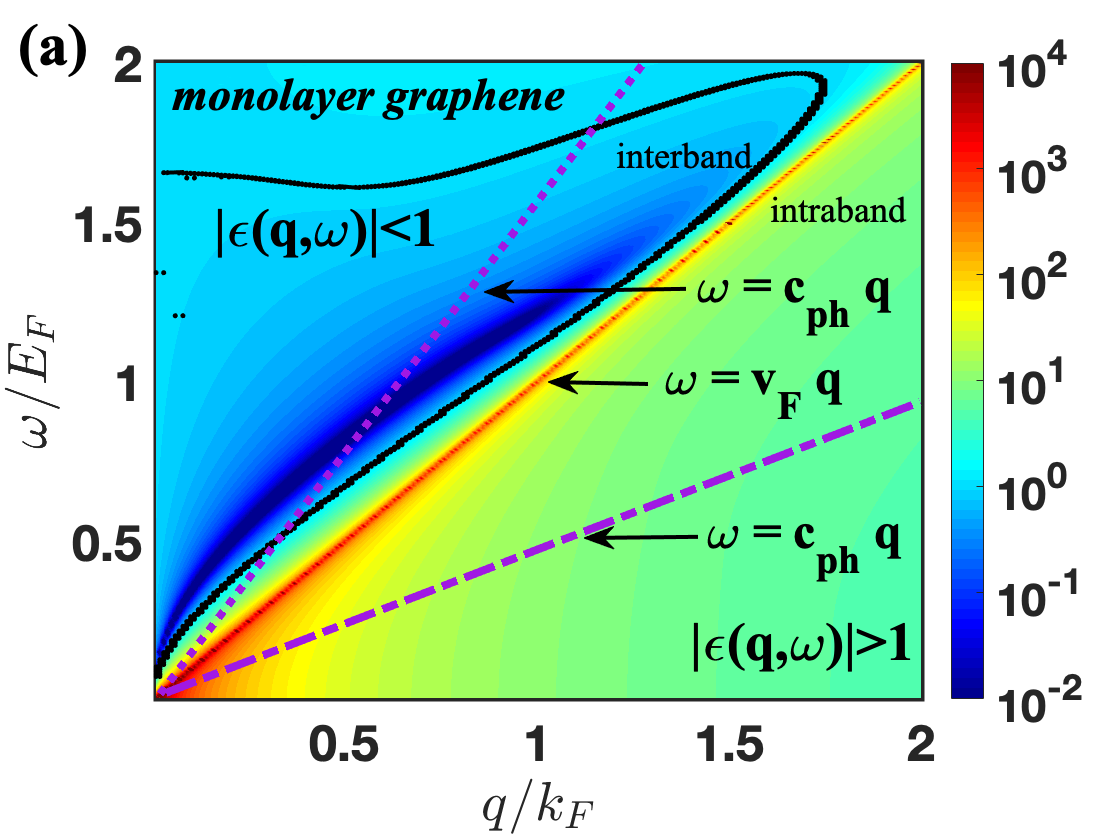}
  \includegraphics[height=!, width=8.5cm]{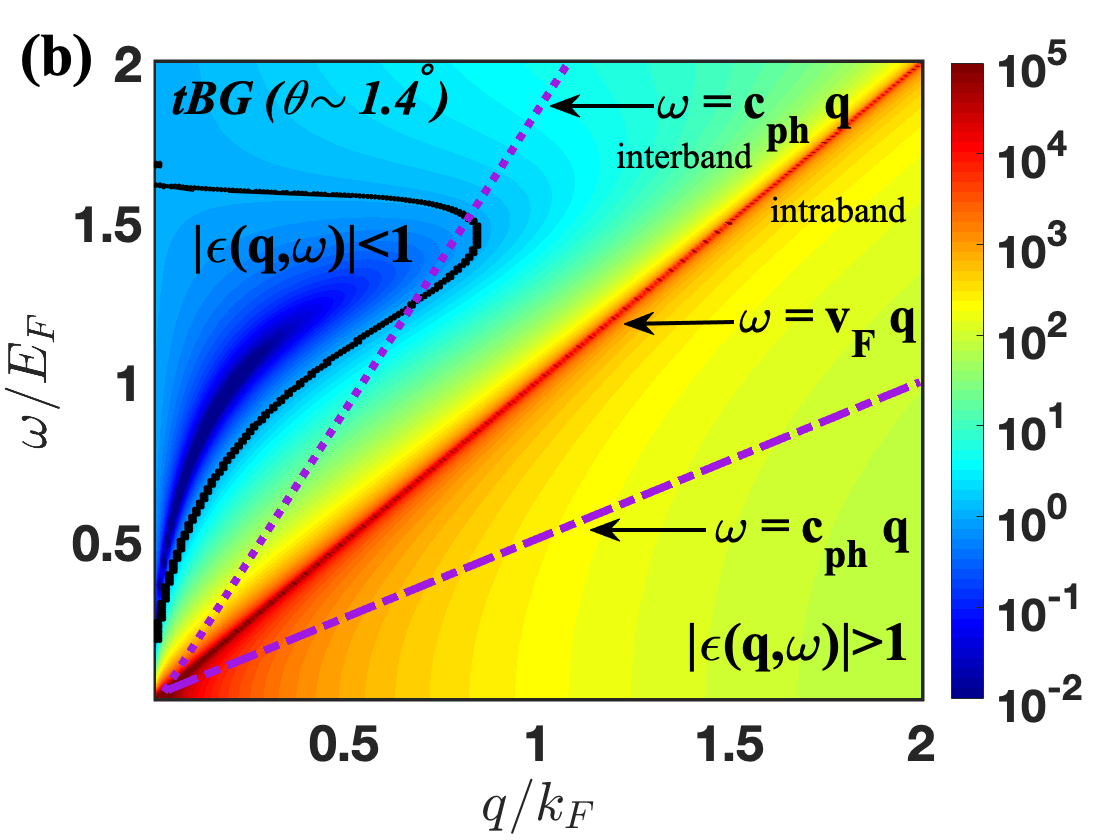}
  \caption{The finite temperature dynamical dielectric function $|\epsilon(q,\omega)|$ in (a) monolayer graphene and (b) tBG at $\theta_M=1.4^{\circ}$ within the Dirac approximation, plotted as a function of $q/k_F$ and $\omega/E_F$ for $T=10$K, $\kappa=4$, and $n=10^{11}$cm$^{-2}$. The regions where $|\epsilon(q,\omega)|>1$  and $|\epsilon(q,\omega)|<1$ are separated by the black curve. The phonon spectrum ($\omega = c_\mathrm{ph}q$) for the intraband process ($c_\mathrm{ph}<v_F$) is indicated by the dashed lines, while the interband process is indicated by the dotted lines.  Notice that for the interband phonons, the spectrum may intersect regions where the dielectric function becomes less than 1, leading to  anti-screening of the electron-phonon vertex at those points. However, this anti-screening region shrinks in tBG due to the reduced Fermi velocity and makes negligible contribution to the electron-phonon scattering rate}
  \label{Fig:epsilon_qomega1}
\end{figure}

\begin{figure}
    \centering
    \includegraphics[width = \columnwidth]{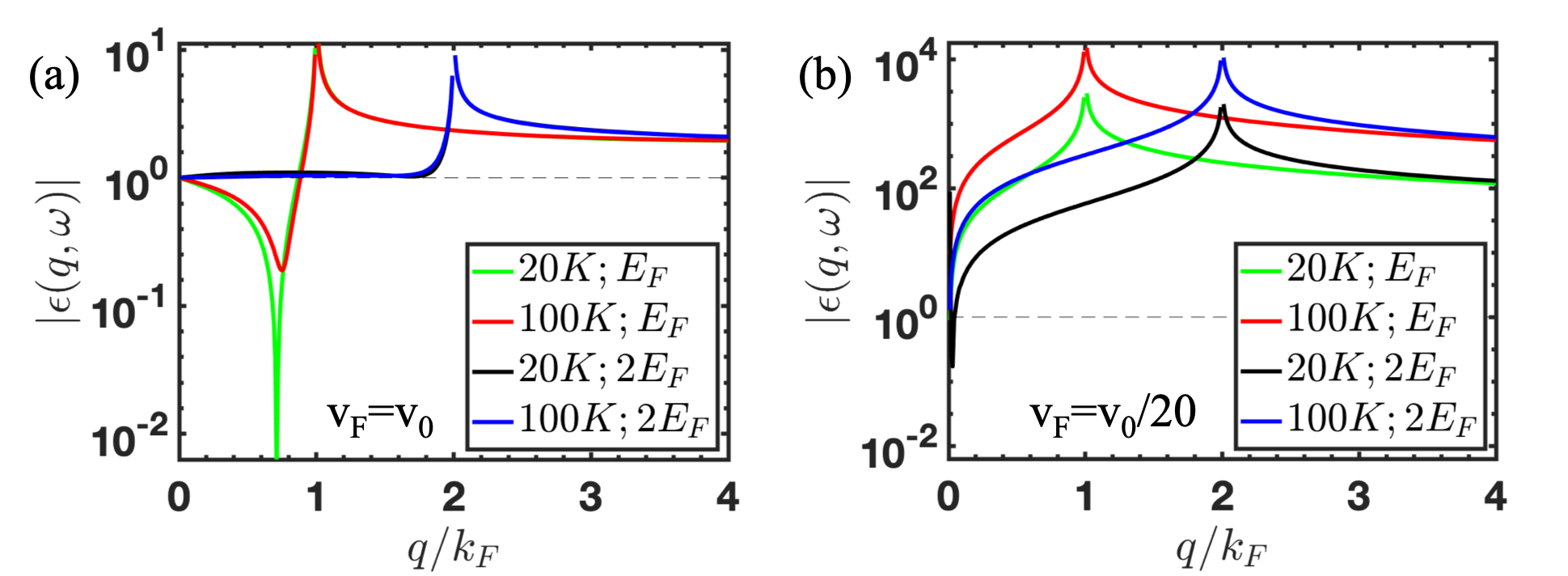}
    \caption{$|\epsilon(q,\omega)|$ for graphene (a) and tBG (b) at different temperatures and frequencies. The suppression of anti-screening and enhancement of screening effects in tBG is observed. Here $v_0$ is the Fermi velocity in monolayer graphene.}
    \label{fig:epscut}
\end{figure}

One might expect that when the plasmon dispersion becomes exactly equal to the phonon dispersion, there will be a divergent contribution leading to overall anti-screening. However it is not the case. The value of the dimensionless wavevector $q/k_F$ where the two dispersions intersect (other than the trivial point $q=0$) is given by $q/k_F = ge^2v_F / 2\kappa \hbar c_\mathrm{ph}^2$. The intersection point decreases at smaller values of $v_F$. Plugging in typical values ($\kappa\sim 4$, $c_\mathrm{ph}\sim 20000$ m/s), and very close to the magic angle ($\theta\sim 1.07^{\circ}$), we find $q/k_F\sim 25$, which is way beyond the interband kinematics regime (As Eq.~\ref{Eqn_tau_inter_full} suggests that the scattering wavevector $q/k\ll 1$ in the kinematically allowed interband regime when $v_F\ll c_\mathrm{ph}$). Thus the possibility of an overall anti-screening is ruled out. In fact, if anti-screening was dominant, it could likely explain the large observed values of deformation potential extracted from the experiments~\cite{wu2018phonon}.
 \begin{figure}[h]
      \centering
     \includegraphics[height=!,width=8.7cm]{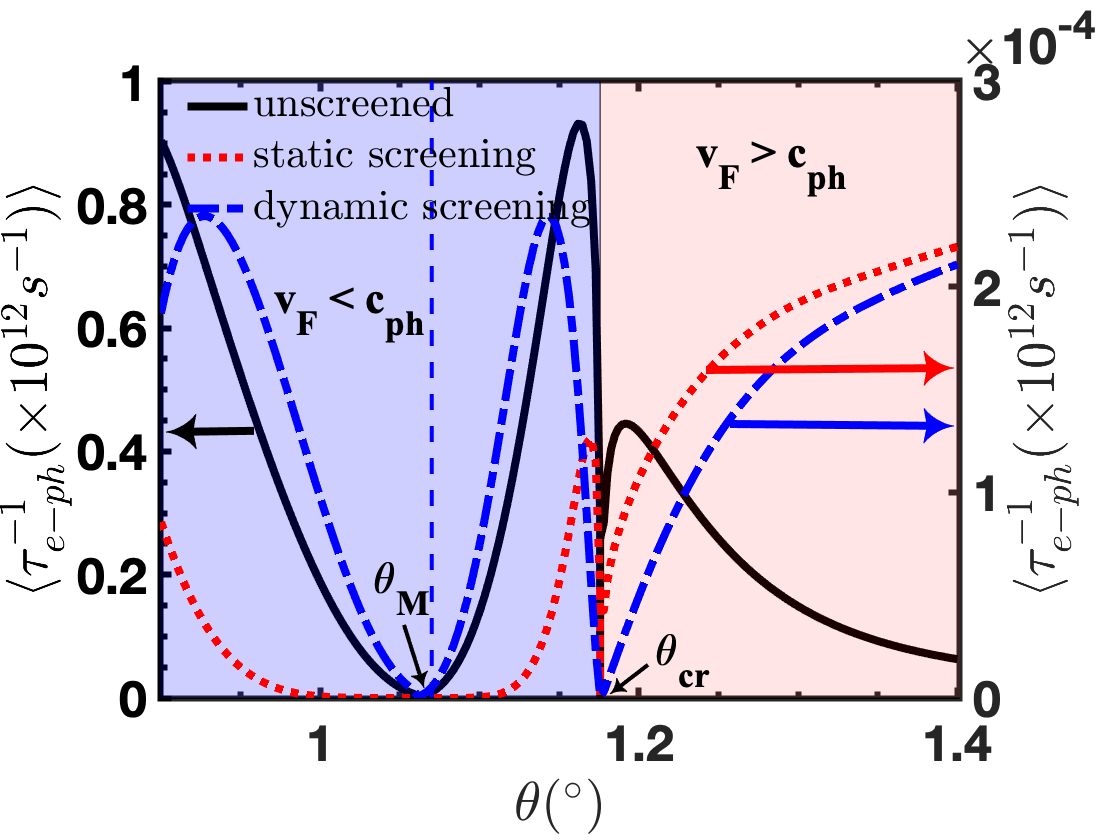}
      \caption{A comparison of energy averaged electron-phonon scattering rates (unscreened vs. static screening vs. dynamic screening) in tBG as a function of twist angle for $T=10$K, $n=5\times 10^{10}$cm$^{-2}$. The shaded red (blue) regions indicate regions when $v_F>c_\mathrm{ph}$ ($v_F<c_\mathrm{ph}$) and the dip occurs at $\theta_\mathrm{cr}$ when $v_F = c_\mathrm{ph}$. The unscreened scattering rate is at least three orders of magnitude larger than the screened version. Further, static screening is qualitatively inaccurate in the interband scattering regime ($c_\mathrm{ph}>v_F$).}
      \label{Fig:Taucomparison}
  \end{figure}
We also explicitly calculate the net effect of screening and anti-screening om the scattering rates. 
We compare the electron-phonon scattering rates for unscreened, static screening, and dynamic screening in tBG as a function of twist angle focusing on the regime close to the magic angle and the critical angle (see Fig.~\ref{Fig:Taucomparison}). Static screening is a good approximation for $\theta>\theta_\mathrm{cr}$ though it could be quantitatively inaccurate close to $\theta_\mathrm{cr}$. However static screening is qualitatively inaccurate when $\theta_M<\theta<\theta_\mathrm{cr}$. Therefore, in order to correctly describe screening properties of scalar phonon modes one must take into account the full frequency dependent dynamical dielectric function. Nevertheless, the unscreened scattering rate is at least three orders of magnitude higher than the dynamically screened scattering rate, even when $\theta_M<\theta<\theta_\mathrm{cr}$. The anti-screening of vertex for few values of momentum $q$ is compensated by the large screening at other values of $q$. The overall screening, and not antiscreening of the electron-phonon coupling, is attributed to the shrinking of the $|\epsilon(q,\omega)|$ phase space in Fig.~\ref{Fig:epsilon_qomega1} for small values of $v_F$. The comparison of resistivities also shows the same pattern as the scattering rates. This explicitly shows that only the gauge phonon modes are relevant and the scalar deformation potential is irrelevant.

\section{Enhancement of antisymmetric gauge phonon mode}
\label{sec:gauge}
\begin{figure}
    \centering
    \includegraphics[width=0.6\columnwidth]{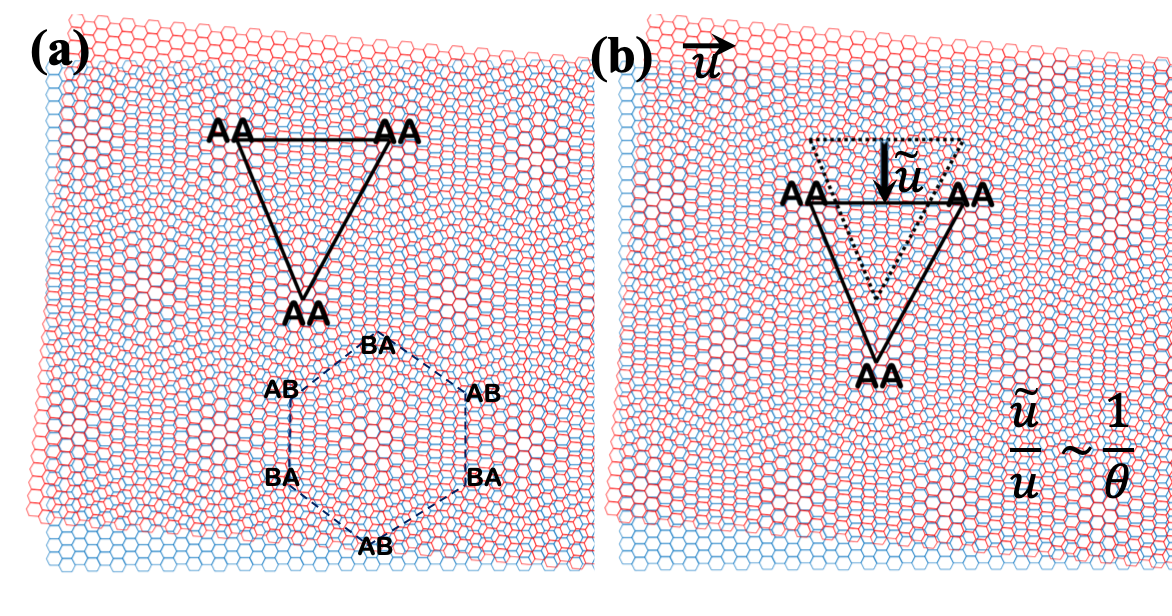}
    \caption{Graphical illustration of the enhancement of antisymmetric gauge phonon mode ($\tilde{\beta}_A$) in tBG.  A small displacement $u$ in one layer causes an enhanced displacement ${\tilde u} \sim 1/\theta$ in the moir\'e superlattice.  (a) superlattice of AA sites (solid lines), and also AB-BA sites (dashed lines) in twisted bilayer graphene, (b) a tiny displacement in the $x$ direction (indicated by horizontal arrow) of the top layer (red) results in a large displacement of the AA sites in the $y$ direction (indicated by vertical arrow), while preserving the area of the superlattice. The dotted and the solid black lines indicate undisplaced and displaced lattice respectively.}
    \label{Fig_tBG_displacement}
\end{figure}

When two layers of graphene are stacked on top of each other, there are four in-plane interlayer acoustic modes (LA1, LA2, TA1, TA2) \cite{cocemasov2013phonons}. Two of these modes are layer symmetric, i.e both layers move in the same direction and the other two are layer anti-symmetric. In this section we calculate the effect of these modes on the electron-phonon coupling of the superlattice.  We can map the antisymmetric modes of tBG on to an effective monolayer by considering the superlattice of AA stacking centers, which form a triangular lattice (see Fig.~\ref{Fig_tBG_displacement}). The AB and BA stacking centers form a dual hexagon superlattice (Fig.~\ref{Fig_tBG_displacement}).
Since within the Dirac model we treat the superlattice as graphene with a renormalized Fermi velocity we can express the electron-phonon coupling matrix for the superlattice as \cite{sohier_phonon-limited_2014,suzuura2002phonons} :

 \begin{equation}
     H_{\text{e-ph}} = \begin{pmatrix}
     D^{'}_A(\tilde{u}_{xx}+\tilde{u}_{yy}) & \beta^{'}_A (\tilde{u}_{xx}-\tilde{u}_{yy}-i(\tilde{u}_{xy}+\tilde{u}_{yx}))\\
     \beta^{'}_A(\tilde{u}_{xx}-\tilde{u}_{yy}+i(\tilde{u}_{xy}+\tilde{u}_{yx})) & D^{'}_A (\tilde{u}_{xx}+\tilde{u}_{yy})
     \end{pmatrix}
     \label{Eq:e-phHamil}
 \end{equation}

 where $D^{'}_A$ and $\beta^{'}_A$ are the deformation potential constant and the gauge field coupling constant of the superlattice respectively. We can write $\beta^{'}_A = \beta_A v_F/v_0$, where $\beta_A$ is the coupling constant for monolayer graphene, since the gauge-field coupling constant is proportional to the Fermi velocity \cite{sohier_phonon-limited_2014}. This factor causes a suppression since there is a reduction in Fermi velocity in tBG near magic angle. $ \tilde{u}_{ij} = \partial \tilde{u}_{i}/\partial j$
 and $\tilde{\textbf{u}}$ is the displacement of the AA/AB sites of the superlattice. Let us first consider the anti-symmetric modes. A relative displacement of $\textbf{u}$ between the two layers causes the AA sites of the superlattice to be displaced by $|\tilde{\textbf{u}}| = \gamma |\textbf{u}|$ in the perpendicular direction \cite{lian_twisted_2018}, where $\gamma = 1/[2\tan(\theta/2)]$. Hence we get:
 \begin{equation}
 \begin{aligned}
     \tilde{u}_{xx} &= \gamma u_{yx}, \quad
     \tilde{u}_{yy} = - \gamma u_{xy}\\ 
     \tilde{u}_{xy} &= \gamma u_{yy},\quad
     \tilde{u}_{yx} = - \gamma u_{xx} 
 \end{aligned}
 \end{equation}
 On substituting these results into Eq.~\ref{Eq:e-phHamil} and using the symmetric property of the in plane strain tensor ($u_{xy} = u_{yx}$) we can write and effective electron-phonon coupling Hamiltonian in terms of the monolayer strain tensor as

 \begin{equation}
     H^A_{\text{e-ph}} = \begin{pmatrix}
     0 & \tilde{\beta}_A(2u_{xy}+i(u_{xx}-u_{yy}))\\
     \tilde{\beta}_A(2u_{xy}-i(u_{xx}-u_{yx})) & 0
     \end{pmatrix}
     \label{Eq:e-phHamil2}
 \end{equation}

\noindent where the superscript $A$ indicates the antisymmetric phonon mode contribution, $\tilde{\beta}_A = \gamma \beta_A (v_F/v_0)$, $\tilde{D}_A = 0$ are the effective gauge field coupling constant and the effective deformation potential constant respectively.  Hence we see that antisymmetric phonon modes cause a large enhancement in the electron-gauge phonon coupling at low angles which counters the reduction in Fermi velocity. We also find that there is no scalar field (diagonal) contribution from the antisymmetric modes as expected because these modes are area preserving.
 The symmetric phonon modes do not have a similar enhancement and they resemble the usual acoustic phonon contribution in graphene, hence
 \cite{koshinoyoungwoo,lian_twisted_2018}.

 \begin{equation}
     H^S_{\text{e-ph}} = \begin{pmatrix}
     D^{'}_A (u_{xx}+u_{yy}) & \beta^{'}_A(u_{xx}-u_{yy}-2iu_{xy})\\
     \beta^{'}_A(u_{xx}-u_{yy}+2iu_{xy}) & D^{'}_A (u_{xx}+u_{yy})
     \end{pmatrix}
     \label{Eq:e-phHamil4}
 \end{equation}

 These symmetric modes have a scalar and a vector field contribution.
 Hence the total electron-phonon coupling Hamiltonian can be written as 
 \begin{equation}
     H_{\text{e-ph}} = H^A_{\text{e-ph}} + H^S_{\text{e-ph}}
 \end{equation}
 At small angles the gauge phonon contribution comes mainly from $H^A_{\text{e-ph}}$ since $\gamma \sim 1/\theta \gg 1$. Hence, gauge phonons become the dominant mechanism for transport at low angles \cite{yudhistira2019gauge}.In monolayer graphene the deformation potential constant is and order of magnitude higher than the gauge field coupling constant, however in tBG, because of large screening effects and no enhancement at small angles the deformation potential contribution becomes irrelevant.
 
 We see the dominance of the geometrically enhanced antisymmetric gauge phonon modes clearly in Fig \ref{Fig_tauinv_vs_theta_threephonons} where we plot the electron-phonon scattering rates for the antisymmetric gauge phonon contribution, the gauge contribution of the symmetric phonon mode and the dynamically screened deformation potential contribution of the symmetric phonon mode. At large angles the first two contributions are similar, however at smaller angles, especially close to the magic angle, because  $\gamma \sim 1/\theta \gg 1$, the antisymmetric gauge phonon mode dominates transport. The bare (unscreened) deformation potential has no fermi velocity renormalization effect neither does it have a geometric enhancement and hence is similar to that of monolayer graphene ($D'_{A} \approx D_A$) \cite{sohier_phonon-limited_2014,wu2018phonon}. Moreover, due to the large screening effect, the deformation potential contribution is irrelevant at all angles.  We note that the validity of these arguments hold only when $\tilde{a}\ll L$, where $L$ is the dimension of the sample, and $\tilde{a} = a_0/[2\sin(\theta/2)]$ is the lattice vector of the moir\'e superlattice, where $a_0=2.46$~\AA~is the graphene lattice constant.  For $L$ of the order of microns, we must have $\theta\gg 0.02^{\circ}$ for the formalism to be valid. We therefore expect that the divergence of $\tilde{\beta}_A$ as $\theta\rightarrow 0$ predicted in this model is unphysical when the moir\'{e}  period becomes comparable to the sample size.
 
 \begin{figure}
    \centering
    \includegraphics[scale=0.22]{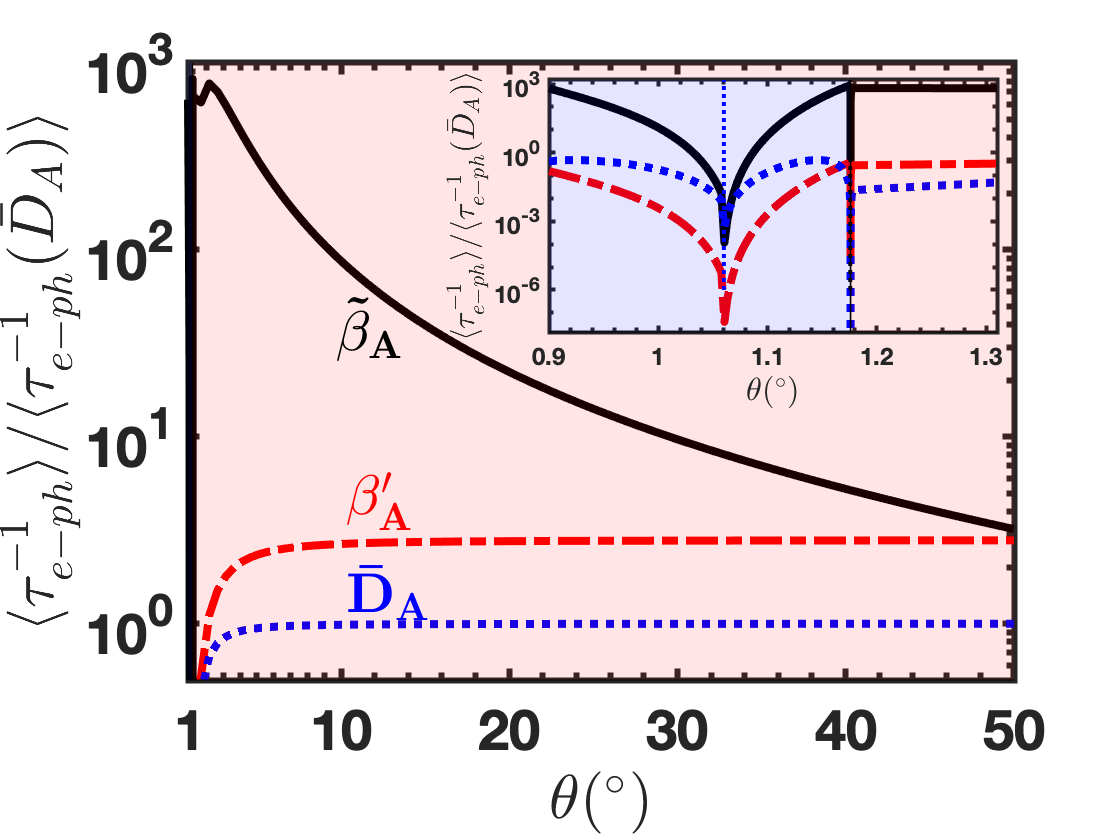}
    \caption{Comparison of electron-phonon scattering rates for the antisymmetric gauge mode ($\tilde{\beta}_A$), symmetric gauge mode ($\beta'_A$), and the dynamically screened scalar (deformation) potential ($\bar{D}_A$). The inset shows the behaviour near magic angle. The scattering rates are normalized with respect to the screened monolayer deformation value $\bar{D}_A$.  The two gauge phonon modes $\tilde{\beta}_A$ and $\beta'_A$ are comparable to each other at large angles; however, near $\theta_M$, $\tilde{\beta}_A$ dominates by several orders of magnitude.}
    \label{Fig_tauinv_vs_theta_threephonons}
\end{figure}

 \section{Boltzmann transport}\label{sec:BT}

\subsection{General formalism}

Carrier current is created by applying an electric field $\mathbf{E}$, which has the effect of changing the electronic distribution $f_{\mathbf{k},\lambda}$ from the Fermi-Dirac distribution $f^0_{\mathbf{k}\lambda}$. Up to linear order in response, the distribution $f_{\mathbf{k}\lambda}$ can be expressed as $f_{\mathbf{k},\lambda} = f^0_{\mathbf{k},\lambda} + h_{\mathbf{k},\lambda}$. The change in the distribution function $h_{\mathbf{k},\lambda}$ can be evaluated within the Boltzmann transport formalism. The change in the distribution function due to the electric field is compensated by the collision integral $\mathrm{St}[f
_{\mathbf{k},\lambda}]$, which describes the rate of change in the occupation of the electronic states due to scattering. 
\begin{align}
     -\frac{e\mathbf{E}}{\hbar}\nabla_\mathbf{k} f_{\mathbf{k},\lambda}=\mathrm{St}[f
_{\mathbf{k},\lambda}] \label{D1}
\end{align}
The collision integral can be written as 
\begin{align}
    \mathrm{St}\left[f_{\mathbf{k},\lambda}\right]=\sum\limits _{\mathbf{k}',\lambda',\nu}P_{\mathbf{k}'\mathbf{k},\nu}^{\lambda'\lambda}f_{\mathbf{k}',\lambda'}(1-f_{\mathbf{k},\lambda})-P_{\mathbf{k}\mathbf{k}',\nu}^{\lambda\lambda'}f_{\mathbf{k},\lambda}(1-f_{\mathbf{k}',\lambda'}),
    \label{Eq_Boltzeqn2}
\end{align}
where $P^{\lambda'\lambda}_{\mathbf{k}'\mathbf{k},\nu}$ is the scattering probability from state $|\lambda', \mathbf{k}'\rangle$ to $|\lambda, \mathbf{k}\rangle$ within phonon branch $\nu$ (TA or LA), which is given by
\begin{align}
    P_{\mathbf{k},\mathbf{k}+\mathbf{q},\nu}^{\lambda\lambda'}&=\frac{2\pi}{\hbar}\left|g_{\mathbf{k},\mathbf{k}+\mathbf{q},\nu}^{\lambda,\lambda'}\right|^{2}\left[n_{\mathbf{q},\nu}\delta(\varepsilon_{\mathbf{k}+\mathbf{q},\lambda'}-\varepsilon_{\mathbf{k},\lambda}-\hbar\omega_{\mathbf{q},\nu})\nonumber\right.\\&\quad\left.+(1+n_{\mathbf{q},\nu})\delta(\varepsilon_{\mathbf{k}+\mathbf{q},\lambda'}-\varepsilon_{\mathbf{k},\lambda}+\hbar\omega_{\mathbf{q},\nu})\right],
\end{align}
where the two terms account for absorption and emission of phonons, $n_{\mathbf{q},\nu}$ is the Bose-Einstein distribution function describing the phonon population, $\hbar \omega_{\mathbf{q},\nu} = \hbar c_\nu q$ is the phonon energy, and $g_{\mathbf{k},\mathbf{k}+\mathbf{q},\nu}^{\lambda,\lambda'}$ is the electron-phonon coupling, which can be expressed as 
\begin{align}
    g_{\mathbf{k},\mathbf{k}+\mathbf{q},\nu}^{\lambda,\lambda'} = \sqrt{\frac{\hbar}{2 A \mu_s \omega_{\mathbf{q},\nu}}} M_{\mathbf{k},\mathbf{k}+\mathbf{q}}^{\lambda,\lambda'}
    \label{Eq_define_g}
\end{align}
where $A$ is the area of the graphene layer, $\rho$ is the mass density, and $M_{\mathbf{k},\mathbf{k}+\mathbf{q}}^{\lambda,\lambda'}$ is the matrix element for scattering between initial and final states, which is given by 
\begin{align}
    M_{\mathbf{k},\mathbf{k}+\mathbf{q}}^{\lambda,\lambda'} = \zeta q \left(\frac{1+\lambda\lambda' \cos \theta_{\mathbf{k},\mathbf{k}+\mathbf{q}}}{2}\right)^{1/2},
    \label{Eq_define_M}
\end{align}
where $\zeta$ is the effective deformation potential and the cosine factor comes from the the chirality of the wavefunctions. The effective deformation potential $\zeta$ stands for either the effective scalar potential $\tilde{D}_A$ or twice the effective gauge potential $2\tilde{\beta}_A$ (see Appendix.~\ref{sec:gauge}). Using the detailed balance condition 
\begin{align}
    P_{\mathbf{k}'\mathbf{k}}^{\lambda'\lambda}f_{\mathbf{k}',\lambda'}^{0}(1-f_{\mathbf{k},\lambda}^{0})=P_{\mathbf{k}\mathbf{k}'}^{\lambda\lambda'}f_{\mathbf{k},\lambda}^{0}(1-f_{\mathbf{k}',\lambda'}^{0}),
\end{align}
one can verify that the the ansatz 
\begin{align}
    h_{\mathbf{k},\lambda}=neE\tau_{\mathbf{k},\lambda}\cos\theta_{\mathbf{k}}v_{F}\frac{\partial f_{\mathbf{k},\lambda}^{0}}{\partial\varepsilon_{\mathbf{k},\lambda}}
\end{align}
solves the Boltzmann equation (Eq.~\ref{D1}), where $\varepsilon_{\mathbf{k},\lambda}$ is the energy dispersion. Furthermore, we assume $\tau_{\mathbf{k'},\lambda'}\approx\tau_{\mathbf{k},\lambda}$, which is a reasonable assumption for impurity assisted electron-phonon scattering~\cite{sohier_phonon-limited_2014}. The scattering time $\tau_{\mathbf{k},\lambda}$ can be solved for as
\begin{align}
    \frac{1}{\tau_{\mathbf{k},\lambda}} = \sum_{\substack{\lambda',\mathbf{k}'\\
\nu=\mathrm{LA,TA}}} (1-\lambda\lambda'\cos \theta_{\mathbf{k}\mathbf{k}'})\frac{1-f_{\mathbf{k}',\lambda'}^{0}}{1-f_{\mathbf{k},\lambda}^{0}} P^{\lambda\lambda'}_{\mathbf{k}\mathbf{k}',\nu},
    \label{Eq:elecphonon1}
\end{align}
where $\theta_{\mathbf{k}\mathbf{k}'}$ is the scattering angle between the initial and final states. The case of $\lambda=\lambda'$ and $\lambda\neq \lambda'$ corresponds to intraband and interband scattering respectively. The momentum dependence of the scattering time can be shown to enter only through the energy dispersion $\epsilon_{\mathbf{k},\lambda} = \lambda\hbar v_F k$ i.e. $\tau^{e-\mathrm{ph}}_{\mathbf{k},\lambda} = \tau^{e-\mathrm{ph}}(\varepsilon_{\mathbf{k},\lambda})$.

Using the fact that 
\begin{align}
    &n_{\mathbf{q},\nu}\frac{1-f_{\mathbf{k}',\lambda'}^{0}}{1-f_{\mathbf{k},\lambda}^{0}} = f_{\mathbf{k}',\lambda'}^{0} + n_{\mathbf{q},\nu},\\
    &(1+n_{\mathbf{q},\nu}) \frac{1-f_{\mathbf{k}',\lambda'}^{0}}{1-f_{\mathbf{k},\lambda}^{0}} = -f_{\mathbf{k}',\lambda'}^{0} + n_{\mathbf{q},\nu}+1,
\end{align}
we find that the scattering rates take the form
\begin{align}
    \frac{1}{\tau_{\mathbf{k},\lambda}^{a}}&=\sum_{\lambda',\nu=\mathrm{TA},\mathrm{LA}}\int\frac{d\mathbf{q}}{(2\pi)}\frac{|M_{\mathbf{k,\mathbf{k}+q}}^{\lambda\lambda'}|^{2}}{2\rho\omega_{\mathbf{q},\nu}}\left(1+\lambda\lambda'\cos\theta_{\mathbf{k},\mathbf{k}+\mathbf{q}}\right)\nonumber\\&\qquad\left(f_{\mathbf{k}',\lambda'}^{0}+n_{\mathbf{q},\nu}\right)\delta\left(\varepsilon_{\mathbf{k}+\mathbf{q},\lambda'}-\varepsilon_{\mathbf{k},\lambda}-\hbar\omega_{\mathbf{q},\nu}\right)
\end{align}
\begin{align}
    \frac{1}{\tau_{\mathbf{k},\lambda}^{e}}&=\sum_{\lambda',\nu=\mathrm{TA},\mathrm{LA}}\int\frac{d\mathbf{q}}{(2\pi)}\frac{|M_{\mathbf{k,\mathbf{k}+q}}^{\lambda\lambda'}|^{2}}{2\rho\omega_{\mathbf{q},\nu}}(1+\lambda\lambda'\cos\theta_{\mathbf{k},\mathbf{k}+\mathbf{q}})\nonumber\\&\qquad\left(1-f_{\mathbf{k}',\lambda'}^{0}+n_{\mathbf{q},\nu}\right)\delta\left(\varepsilon_{\mathbf{k}+\mathbf{q},\lambda'}-\varepsilon_{\mathbf{k},\lambda}+\hbar\omega_{\mathbf{q},\nu}\right)
\end{align}
where the subscript $a$ ($e$) indicate phonon absorption (emission) respectively and the total rate is the sum of the two. The delta functions can be simplified, and the interband scattering rates can be evaluated to be

 \begin{align}
 &\frac{1}{\tau^{\text{inter}}_{a}(\mathbf{k})} = \sum_{\nu=\mathrm{TA},\mathrm{LA}}\frac{\zeta(\theta)^{2}}{2\hbar\mu_s c_{\nu}}\int_{2k/(z+1)}^{2k/(z-1)}\frac{dq}{2\pi}q^{3}\frac{\sqrt{1-s_{-,\nu}^{2}}}{k\sqrt{k^{2}+q^{2}+2kqs_{-,\nu}}}\left[f_{0}(\hbar v_{F}\sqrt{k^{2}+q^{2}+2kqs_{-,\nu}})+n_{\mathbf{q},\nu}\right] \nonumber\\
	&\frac{1}{\tau^{\text{inter}}_{e}(\mathbf{k})} = \sum_{\nu=\mathrm{TA},\mathrm{LA}}\frac{\zeta(\theta)^{2}}{2\hbar\mu_s c_{\nu}}\int_{2k/(z+1)}^{2k/(z-1)}\frac{dq}{2\pi}q^{3}\frac{\sqrt{1-s_{-,\nu}^{2}}}{k\sqrt{k^{2}+q^{2}+2kqs_{-,\nu}}}\left[1-f_{0}(-\hbar v_{F}\sqrt{k^{2}+q^{2}+2kqs_{-,\nu}})+n_{\mathbf{q},\nu}\right]\nonumber\\
	\label{Eqn_tau_inter_full}
	\end{align}

where $ s_{-,\nu}=(q/2k) (z_\nu^2 - 1) - z_\nu$, $z_\nu = c_\nu/v_F$, $f_0(x) = 1/\{\exp[(x-\mu)/k_{B}T]+1\}$.

For intraband scattering, in the limit of $c_\nu\ll v_F$ we have 
\begin{align}
\frac{1}{\tau^{\text{intra}}_{a}(\mathbf{k})} = \sum_{\nu=\mathrm{TA},\mathrm{LA}}\frac{\zeta^2}{2\hbar \mu_s c_\nu v_F} \int_{0}^{2k} \frac{dq}{2\pi} \frac{q^3}{k^2}\sqrt{1-(q/2k)^2} (f_{\epsilon_\mathbf{k}} + n_{\mathbf{q},\nu}),
\end{align}
while
\begin{align}
\frac{1}{\tau^{\text{intra}}_{e}(\mathbf{k})} = \sum_{\nu=\mathrm{TA},\mathrm{LA}}\frac{\zeta^2}{2\hbar \mu_s c_\nu v_F} \int_{0}^{2k} \frac{dq}{2\pi} \frac{q^3}{k^2}\sqrt{1-(q/2k)^2} (1-f_{\epsilon_\mathbf{k}} + n_{\mathbf{q},\nu}),
\end{align}
The total scattering rate is 
\begin{align}
\frac{1}{\tau^{\text{intra}} (\mathbf{k})} = \sum_{\nu=\mathrm{TA},\mathrm{LA}}\frac{\zeta^2}{2\hbar \mu_s c_\nu v_F} \int_{0}^{2k} \frac{dq}{2\pi} \frac{q^3 \sqrt{1-(q/2k)^2}}{k^2} (1 + 2 n_{\mathbf{q},\nu}),
\end{align}

The resistivity $\rho_{e-\mathrm{ph}}$ is obtained from the scattering time $\tau^{e-\mathrm{ph}} (\varepsilon)$ by the energy average
\begin{equation}
\frac{1}{\rho_{e-ph}} = e^2 \int d\varepsilon N_D(\varepsilon) \frac{v_F^2}{2}\tau^{e-\mathrm{ph}}(\varepsilon) \left(-\frac{\partial f^0(\varepsilon)}{\partial \varepsilon}\right),
\label{eq:rho_formula}
\end{equation}
where $N_D$ is density of states.

\subsection{Beyond the Dirac approximation}

In this subsection, we provide a detailed explanation of the effective model in eq.~\ref{CN} of the main text including density of states (DOS), scattering time and resistivity. Without loss of generality, we locate K and K' points along $k_{y}$ axis, i.e. $\Delta K=ik_{\theta}$
Hence
\begin{equation}
   H=-\frac{\hbar v_{F}}{k_{\theta}}\begin{pmatrix}0 &  k^{*2}-\left(-ik_{\theta}/2\right)^{2}\\
k^{2}-\left(ik_{\theta}/2\right)^{2} & 0
\end{pmatrix}, 
\end{equation}
where $k = k_x + ik_y$.
Introducing $\varepsilon_\mathrm{VHS}=(1/4)\hbar v_{F}k_{\theta}$, which corresponds to energy at van Hove singularity, we can write the Hamiltonian as
\[
H=-\varepsilon_\mathrm{VHS}\begin{pmatrix}0 & 4\left(\tilde{k}_{x}-i\tilde{k}_{y}\right)^{2}+1\\
4\left(\tilde{k}_{x}+i\tilde{k}_{y}\right)^{2}+1 & 0
\end{pmatrix},
\]
where $\tilde{\mathbf{k}}=\mathbf{k}/k_{\theta}$.

To circumvent difficulties with anisotropic dispersion, we map this
Hamiltonian to massless Dirac Hamiltonian
 \begin{equation}
     H = \varepsilon_\mathrm{VHS} r 
     \begin{pmatrix}
     0 & e^{-i\phi}\\
     e^{i\phi} & 0
     \end{pmatrix}
     \label{isotropform}
\end{equation}
 where $\varepsilon_\mathrm{VHS} = (1/4) \hbar v_F k_{\theta}$.
 
This can be achieved by using the following variable transformation from ($k_x,k_y$) to $(r,\phi)$

 \begin{align}
k_{x} & =(\gamma/2)k_{\theta}\sqrt{\frac{1}{2}\left[-(1+r\cos\phi)+\sqrt{(1+r\cos\phi)^{2}+(r\sin\phi)^{2}}\right]}\\
k_{y} & =-(\gamma/2)k_{\theta}\mathrm{sgn}\left(\sin\phi\right)\sqrt{\frac{1}{2}\left[1+r\cos\phi+\sqrt{(1+r\cos\phi)^{2}+(r\sin\phi)^{2}}\right]},
\label{vartransform}
\end{align}
where $\gamma=\pm1$ represents each half of the Fermi surface. Note that Fermi surface splitting happens at energy $|E|<\varepsilon_\mathrm{VHS}$. The expression in eq.~\ref{isotropform} resembles that of monolayer graphene and hence we can calculate the resistivity as in monolayer graphene, with the caveat that we also need to introduce the determinant of the Jacobian matrix of the transformation in the integrals. 

The range of the new co-ordinate $\phi$ is $\phi\in[-\pi,\pi].$
In the transformed coordinates, the energy dispersion and the corresponding eigenstates are given by 
\begin{align}
E_{\pm}(r)&=\pm\varepsilon_\mathrm{VHS}r \\
|\pm\rangle&=\frac{1}{\sqrt{2}}\begin{pmatrix}\pm1\\
e^{i\phi}
\end{pmatrix}
\end{align}

Note that in the new coordinates, $r=1$ corresponds to van Hove
singularity in the original coordinates.
The Jacobian $\mathcal{J}$ corresponding to this transformation is
given by
\begin{align}
\mathcal{\mathcal{J}}(r,\phi) & =\begin{vmatrix}\frac{\partial k_{x}}{\partial r} & \frac{\partial k_{x}}{\partial\phi}\nonumber\\
\frac{\partial k_{y}}{\partial r} & \frac{\partial k_{y}}{\partial\phi}
\end{vmatrix}\\
 & =\frac{k_{\theta}^{2}r}{16\sqrt{(1+r\cos\phi)^{2}+(r\sin\phi)^{2}}}
\end{align}

The density of states at energy below the Van Hove singularity $|\varepsilon|<\varepsilon_\mathrm{VHS}$ can be obtained analytically as
\begin{align}
D(\varepsilon) & =\frac{g}{2}\int\frac{d^{2}k}{(2\pi)^{2}}\delta(\varepsilon-\varepsilon_{\mathbf{k},+})\nonumber\\
 & =\frac{g}{2}\sum_{\gamma=\pm1}\int_{0}^{\infty}dr\int_{-\pi}^{\pi}d\phi\frac{\mathcal{J}(r,\phi)}{(2\pi)^{2}}\delta(\varepsilon-\varepsilon_\mathrm{VHS}r)\nonumber\\
 & =\frac{1}{2}\left(\frac{n_{\mathrm{VHS}}}{\varepsilon_\mathrm{VHS}}\right)\left(\frac{|\varepsilon|}{\varepsilon_\mathrm{VHS}}\right)\sum_{z=\pm1}\frac{K\left[4z\frac{|\varepsilon|/\varepsilon_\mathrm{VHS}}{\left(1+z|\varepsilon|/\varepsilon_\mathrm{VHS}\right)^{2}}\right]}{1+z\left(|\varepsilon|/\varepsilon_\mathrm{VHS}\right)},
\end{align}
where $g=8$ is the degeneracy, $K(x)=(\pi/2)\sum_{n=0}^{\infty}[(2n-1)!!/(2n)!!]^{2}x^{2n}$ is the complete elliptic integral of the first kind, and $n_\mathrm{VHS}$ is the density at VHS, given by $n_{\mathrm{VHS}}=[g/(16\pi^{2})]k_{\theta}^{2}$.

In the following paragraph, we provide a detailed calculation of the scattering time from the effective model in described above. The electron-phonon scattering time can be obtained from the standard Boltzmann formula
\begin{align}
\frac{1}{\tau_{\mathbf{k},\lambda}^{(j)}} & =\sum_{\substack{\lambda',\mathbf{k}'\\
\nu=\mathrm{LA,TA}}}P_{\mathbf{k},\mathbf{k}',\nu}^{\lambda,\lambda'}\frac{1-f_{\mathbf{k}',\lambda'}^{0}}{1-f_{\mathbf{k},\lambda}^{0}}\left(1-\frac{v_{\mathbf{k'},\lambda'}^{(j)}}{v_{\mathbf{k},\lambda}^{(j)}}\frac{\tau_{\mathbf{k}',\lambda'}^{(j)}}{\tau_{\mathbf{k},\lambda}^{(j)}}\right)\nonumber\\
 & =A\sum_{\lambda',\nu=\mathrm{TA},\mathrm{LA}}\int\frac{d^{2}\mathbf{k}'}{(2\pi)^{2}}P_{\mathbf{k},\mathbf{k}',\nu}^{\lambda,\lambda'}\frac{1-f_{\mathbf{k}',\lambda'}^{0}}{1-f_{\mathbf{k},\lambda}^{0}}\left(1-\frac{v_{\mathbf{k'},\lambda'}^{(j)}}{v_{\mathbf{k},\lambda}^{(j)}}\frac{\tau_{\mathbf{k}',\lambda'}^{(j)}}{\tau_{\mathbf{k},\lambda}^{(j)}}\right)
\end{align}
which in the new coordinates $\textbf{p}' = (r',\phi')$ can be written as

\begin{equation}
    \frac{1}{\tau_{\mathbf{k},\lambda,\gamma}^{(j)}}=A\sum_{\gamma',\lambda',\nu}\int_{0}^{\infty}dr'\int_{-\pi}^{\pi}d\phi'\frac{\mathcal{J}(r',\phi')}{(2\pi)^{2}}P_{\mathbf{k},\mathbf{k}',\nu}^{\lambda,\lambda';\gamma,\gamma'}\frac{1-f_{\mathbf{k}',\lambda'}^{0}}{1-f_{\mathbf{k},\lambda}^{0}}\left(1-\frac{v_{\mathbf{k'},\lambda',\gamma'}^{(j)}}{v_{\mathbf{k},\lambda,\gamma}^{(j)}}\frac{\tau_{\mathbf{k}',\lambda',\gamma'}^{(j)}}{\tau_{\mathbf{k},\lambda,\gamma}^{(j)}}\right)
\end{equation}

We include only electron-phonon intravalley scattering, i.e.
\begin{equation}
P_{\mathbf{k},\mathbf{k}',\nu}^{\lambda,\lambda';\gamma,\gamma'}=\begin{cases}
P_{\mathbf{k},\mathbf{k}',\nu}^{\lambda,\lambda'} & ;\gamma'=\gamma\\
0 & ;\gamma'\neq\gamma
\end{cases},
\end{equation}
and we make the following approximation
\begin{equation}
    1-\frac{v_{\mathbf{k'},\lambda'}^{(j)}}{v_{\mathbf{k},\lambda}^{(j)}}\frac{\tau_{\mathbf{k}',\lambda'}}{\tau_{\mathbf{k},\lambda}}  \rightarrow1-\cos\phi_{\mathbf{p},\mathbf{p}'}^{\prime},
\end{equation}
which have negligible effect in resistivity calculation.

Hence, we can express the scattering rate as

\begin{equation}
    \frac{1}{\tau}\approx A\sum_{\substack{\lambda'\\ \nu=\mathrm{TA},\mathrm{LA}}}\int_{0}^{\infty}dr'\int_{-\pi}^{\pi}d\phi_{\mathbf{p},\mathbf{p}'}^{\prime}\frac{\mathcal{J}(r',\phi_{\mathbf{p},\mathbf{p}'}^{\prime})}{(2\pi)^{2}}P_{\mathbf{k},\mathbf{k}',\nu}^{\lambda,\lambda'}\frac{1-f_{\mathbf{k}',\lambda'}^{0}}{1-f_{\mathbf{k},\lambda}^{0}}\left(1-\cos\phi_{\mathbf{p},\mathbf{p}'}^{\prime}\right)
\end{equation}

Now we study the separate contributions from intraband and interband scattering.
\subsubsection{Intraband $(v_{F}>c):$}

The intraband scattering rate is
\begin{align}
\frac{1}{\tau} & =\frac{1}{\tau^{(a)}}+\frac{1}{\tau^{(e)}}\\
\frac{1}{\tau^{(a)}}&=\sum_{\nu=\mathrm{TA},\mathrm{LA}}\frac{\tilde{\beta}_{A}}{\pi\mu_{s}\hbar^{2}v_{F}c_\nu}\int_{0}^{2p/\left(1-z_\nu\right)}dq\frac{1}{\sqrt{\left(1+r'\cos\phi_{\mathbf{p},\mathbf{p}+\mathbf{q}}^{\prime\star}\right)^{2}+\left(r'\sin\phi_{\mathbf{p},\mathbf{p}+\mathbf{q}}^{\prime\star}\right)^{2}}}\frac{q^{3}}{p^{2}}\frac{\sqrt{1-\cos^{2}\phi_{\mathbf{p},\mathbf{q}}^{\prime\star}}}{1+z_\nu(q/p)}\left(n_{\mathbf{q},\nu}+f_{\mathbf{p}',\lambda'}^{0}\right)\\
\frac{1}{\tau^{(e)}}& =\sum_{\nu=\mathrm{TA},\mathrm{LA}}\frac{\tilde{\beta}_{A}}{\pi\mu_{s}\hbar^{2}v_{F}c_\nu}\int_{0}^{2p/\left(1+z_\nu\right)}dq\frac{1}{\sqrt{\left(1+r'\cos\phi_{\mathbf{p},\mathbf{p}+\mathbf{q}}^{\prime\star}\right)^{2}+\left(r'\sin\phi_{\mathbf{p},\mathbf{p}+\mathbf{q}}^{\prime\star}\right)^{2}}}\frac{q^{3}}{p^{2}}\frac{\sqrt{1-\cos^{2}\phi_{\mathbf{p},\mathbf{q}}^{\prime\star}}}{1-z_\nu(q/p)}\left(1+n_{\mathbf{q},\nu}-f_{\mathbf{p}',\lambda'}^{0}\right),
\end{align}
where $z_\nu=c_\nu/v_{F}$, $p=(r/4)k_{\theta}$, and the superscript $\star$ means that $\phi_{\mathbf{p},\mathbf{p}+\mathbf{q}}^{\prime\star}$ must satisfy the conservation of energy and momentum, i.e. the solution of delta function. 

Expressing $r'$ and $\phi_{\mathbf{p},\mathbf{p}+\mathbf{q}}^{\prime}$
in terms of $r$ and $\phi_{\mathbf{p},\mathbf{q}}^{\prime}$ i.e.
\begin{align}
r'\cos\phi_{\mathbf{p},\mathbf{p}+\mathbf{q}}^{\prime} & =4\frac{p'}{k_{\theta}}\frac{p+q\cos\phi_{\mathbf{p},\mathbf{q}}^{\prime}}{p'}\nonumber\\
 & =r\left(1+\frac{q}{p}\cos\phi_{\mathbf{p},\mathbf{q}}^{\prime}\right)
\end{align}
and
\begin{align}
r'\left|\sin\phi_{\mathbf{p},\mathbf{p}+\mathbf{q}}^{\prime}\right| & =r'\sqrt{1-\cos^{2}\phi_{\mathbf{p},\mathbf{p}+\mathbf{q}}^{\prime}}\nonumber\\
 & =r'\sqrt{1-\left(\frac{p+q\cos\phi_{\mathbf{p},\mathbf{q}}^{\prime}}{p'}\right)^{2}}\nonumber\\
 & =r'\frac{q}{p'}\left|\sin\phi_{\mathbf{p},\mathbf{q}}^{\prime}\right|\nonumber\\
 & =r\frac{q}{p}\left|\sin\phi_{\mathbf{p},\mathbf{q}}^{\prime}\right|,
\end{align}
we can write the scattering time as

\begin{align}
\frac{1}{\tau^{(a)}} & =\sum_{\nu=\mathrm{TA},\mathrm{LA}}\frac{\tilde{\beta}_{A}}{\pi\mu_{s}\hbar^{2}v_{F}c_\nu}\int_{0}^{2p/\left(1-z_\nu\right)}dq\frac{1}{\sqrt{\left\{ 1+r\left[1+(q/p)s_{+,\nu}\right]\right\} ^{2}+\left[r(q/p)\right]^{2}(1-s_{+,\nu}^{2})}}\frac{q^{3}}{p^{2}}\frac{\sqrt{1-s_{+,\nu}^{2}}}{1+(c_\nu/v_{F})(q/p)}\nonumber\\
& \qquad\qquad\quad\left[\frac{1}{e^{\beta\hbar c_\nu q}-1}+\frac{1}{e^{\beta\left(\hbar v_{F}p+\hbar c_\nu q-\mu\right)}+1}\right]\\
\frac{1}{\tau^{(e)}} & =\sum_{\nu=\mathrm{TA},\mathrm{LA}}\frac{\tilde{\beta}_{A}}{\pi\mu_{s}\hbar^{2}v_{F}c_\nu}\int_{0}^{2p/\left(1+z_\nu\right)}dq\frac{1}{\sqrt{\left\{ 1+r\left[1+(q/p)s_{-,\nu}\right]\right\} ^{2}+\left[r(q/p)\right]^{2}(1-s_{-,\nu}^{2})}}\frac{q^{3}}{p^{2}}\frac{\sqrt{1-s_{-,\nu}^{2}}}{1-(c_\nu/v_{F})(q/p)}\nonumber\\
 & \qquad\qquad\quad\left[1+\frac{1}{e^{\beta\hbar c_\nu q}-1}-\frac{1}{e^{\beta\left(\hbar v_{F}p-\hbar c_\nu q-\mu\right)}+1}\right],
\end{align}
where $z_\nu=c_\nu/v_{F}$, $s_{\pm,\nu}=[q/(2k)](z_\nu^{2}-1)\pm z$ and $p=(r/4)k_{\theta}$. Combining both scattering terms for absorption and emission leads to Eq.~\ref{eq:scattrateephCN} in the main text.

\subsubsection{Interband $(v_{F}<c):$}

Similarly for interband we get the following expression for scattering rate.

\begin{multline}
\frac{1}{\tau} =\sum_{\nu=\mathrm{TA},\mathrm{LA}}\frac{\tilde{\beta}_{A}}{\pi\mu_{s}\hbar^{2}v_{F}c}\int_{2p/(z_\nu+1)}^{2p/(z_\nu-1)}dq\frac{1}{\sqrt{\left\{ 1+r\left[1+(q/p)s_{-,\nu}\right]\right\} ^{2}+\left[r(q/p)\right]^{2}(1-s_{-,\nu}^{2})}}\frac{q^{3}}{p^{2}}\frac{\sqrt{1-s_{-,\nu}^{2}}}{(c_\nu/v_{F})(q/p)-1}\\
\left[\frac{1}{e^{\beta\hbar c_\nu q}-1}+\frac{1}{e^{\beta\left(-\hbar v_{F}p+\hbar c_\nu q-\mu\right)}+1}\right],
\end{multline}
where $z_\nu=c_\nu/v_{F}$, $s_{-,\nu}=[q/(2p)](z_\nu^{2}-1)-z_\nu$.

Finally, resistivity is given by
\begin{align}
\frac{1}{\rho_{ij}} & =e^{2}\frac{g}{2}\int\frac{d^{2}k}{(2\pi)^{2}}v_{\mathbf{k},\lambda}^{(i)}v_{\mathbf{k},\lambda}^{(j)}\tau^{e-\mathrm{ph}}(r)\left(-\frac{\partial f^{0}(\varepsilon)}{\partial\varepsilon}\right)\\
 & =e^{2}g\int_{0}^{\infty}dr\int_{-\pi}^{\pi}d\phi\frac{\mathcal{J}(r,\phi)}{(2\pi)^{2}}v_{\mathbf{k},\lambda}^{(i)}v_{\mathbf{k},\lambda}^{(j)}\tau^{e-\mathrm{ph}}(r)\left(-\frac{\partial f^{0}(\varepsilon)}{\partial\varepsilon}\right),
\end{align}
where $g=8$ (degeneracy) and $v_{\mathbf{k},\lambda}^{(j)}$ is band velocity in $j$ direction.

It can be simplified into
 \begin{align}
\frac{1}{\rho_{\mathrm{Pl}}^{ij}} & =\frac{e^{2}}{h}\frac{1}{C}\left(\frac{\varepsilon_{\mathrm{VHS}}}{k_{B}T}\right)^{2}\nonumber\\
& \qquad\times\frac{1}{\pi}\sum_{\lambda=\pm1}\int_{0}^{\infty}\int_{-\pi}^{\pi}d\phi\frac{\tilde{\mathcal{J}}(r,\phi)\tilde{v}_{\mathbf{k},\lambda}^{(i)}\tilde{v}_{\mathbf{k},\lambda}^{(j)}}{\cosh^{2}\left[\frac{1}{2}\left(\lambda r\frac{\varepsilon_{\mathrm{VHS}}}{k_{B}T}-\tilde{\mu}\right)\right]}\nonumber\\
& =\frac{e^{2}}{h}\frac{1}{C}K_{j}\left(\frac{n}{n_{\mathrm{VHS}}},\frac{k_{B}T}{\varepsilon_{\mathrm{VHS}}}\right),
\end{align}
where the function $K_{j}$ is computed numerically.

The band velocity $v_{\mathbf{k},\lambda}^{(j)}=(1/\hbar)(\partial\varepsilon_{\mathbf{k},\lambda}/\partial k_j)$ in this anisotropic model in $x$ and $y$ direction is given by
\begin{align}
v_{\mathbf{k},\lambda}^{(x)} & =\lambda\gamma v_{F}\frac{u(r,\phi)+1}{r}\sqrt{\frac{1}{2}\left[u(r,\phi)-(1+r\cos\phi)\right]}\\
v_{\mathbf{k},\lambda}^{(y)} & =-\lambda\gamma v_{F}\mathrm{sgn}\left(\sin\phi\right)\frac{u(r,\phi)-1}{r}\sqrt{\frac{1}{2}\left[u(r,\phi)+1+r\cos\phi\right]},
\end{align}
where $u(r,\phi)\equiv\sqrt{(1+r\cos\phi)^{2}+(r\sin\phi)^{2}}$.

\section{Discussion about available experiments}
\label{sec:expt}

\begin{figure}
\includegraphics[scale=0.45]{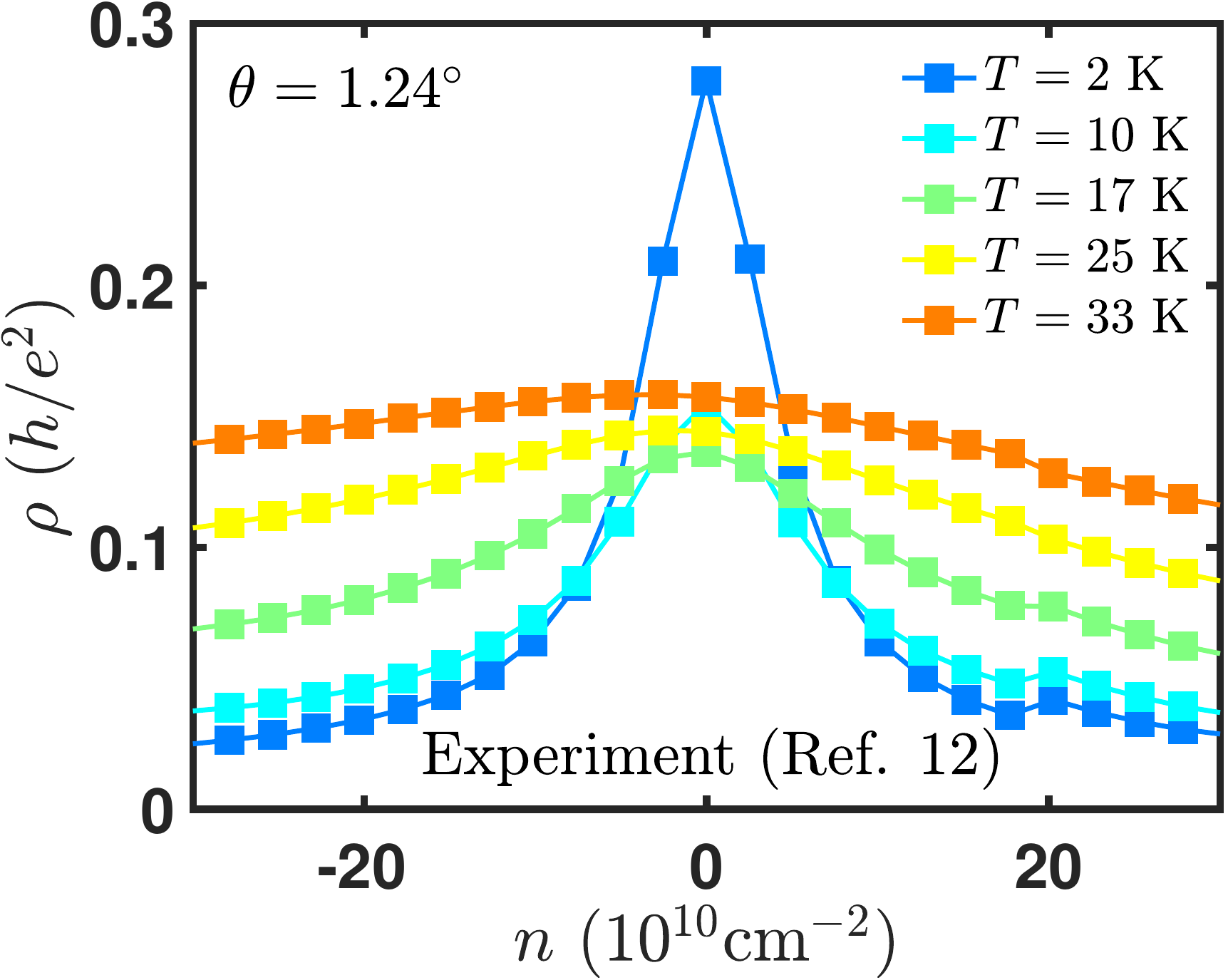}
\llap{\parbox[b]{6.4in}{\Large{(a)}\\\rule{0ex}{2.25in}}}
\includegraphics[scale=0.45]{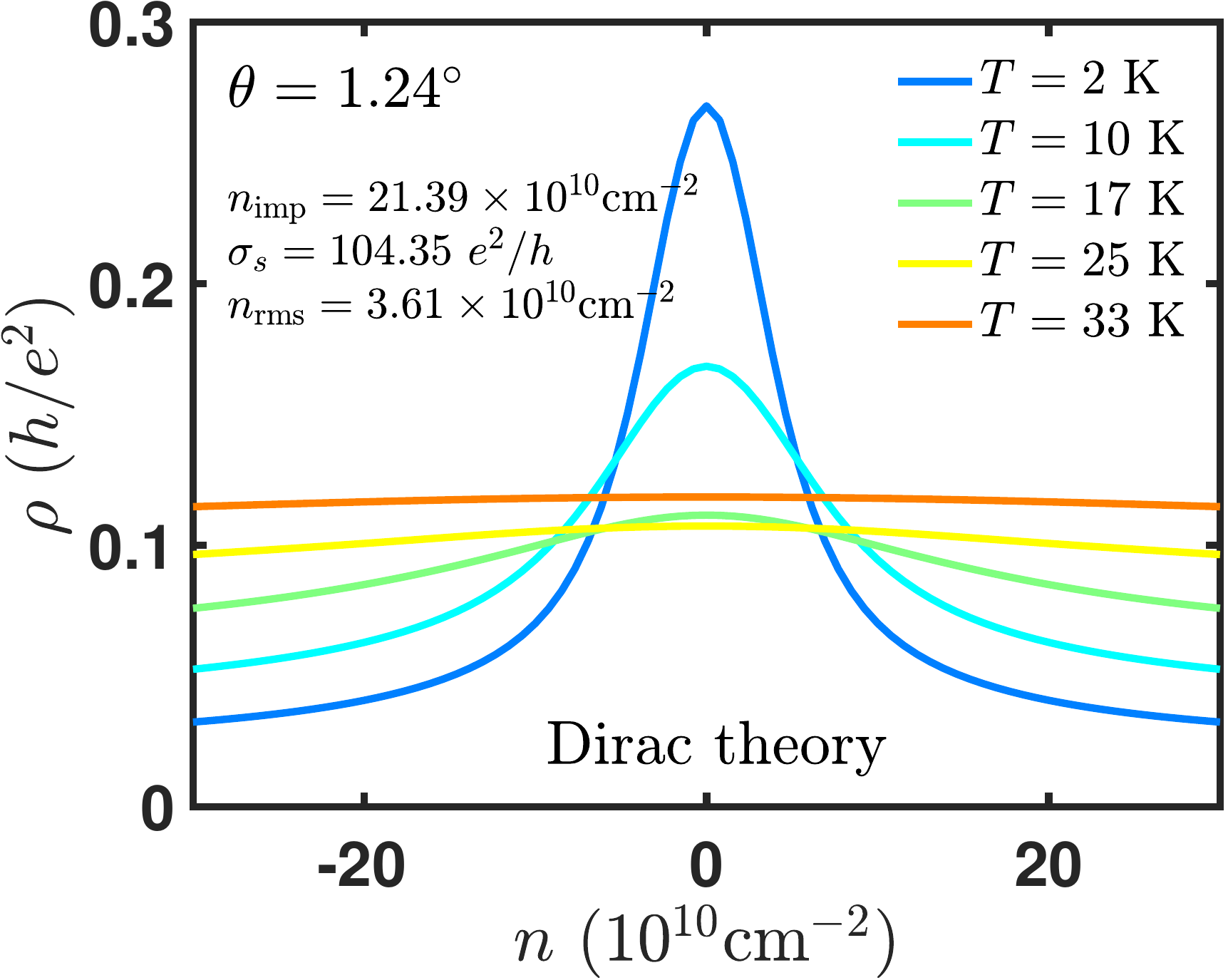}
\llap{\parbox[b]{6.4in}{\Large{(b)}\\\rule{0ex}{2.25in}}}
\caption{The weak density dependence at high temperature and strong density dependence at low temperature of the resistivity close to charge neutrality seen in experiment \cite{polshyn_2019_large} is well captured within the Dirac theory of electron-(gauge) phonon scattering \cite{yudhistira2019gauge}. Experimental data (top squares) for resistivity at $\theta=1.24^\circ$  vs density at various temperatures (a) compared to the Dirac theory (bottom solid lines) of electron-phonon scattering (b).  \label{Fig:rho_exp_Dirac_vsn}}
\end{figure}

\begin{figure}
\includegraphics[scale=0.45]{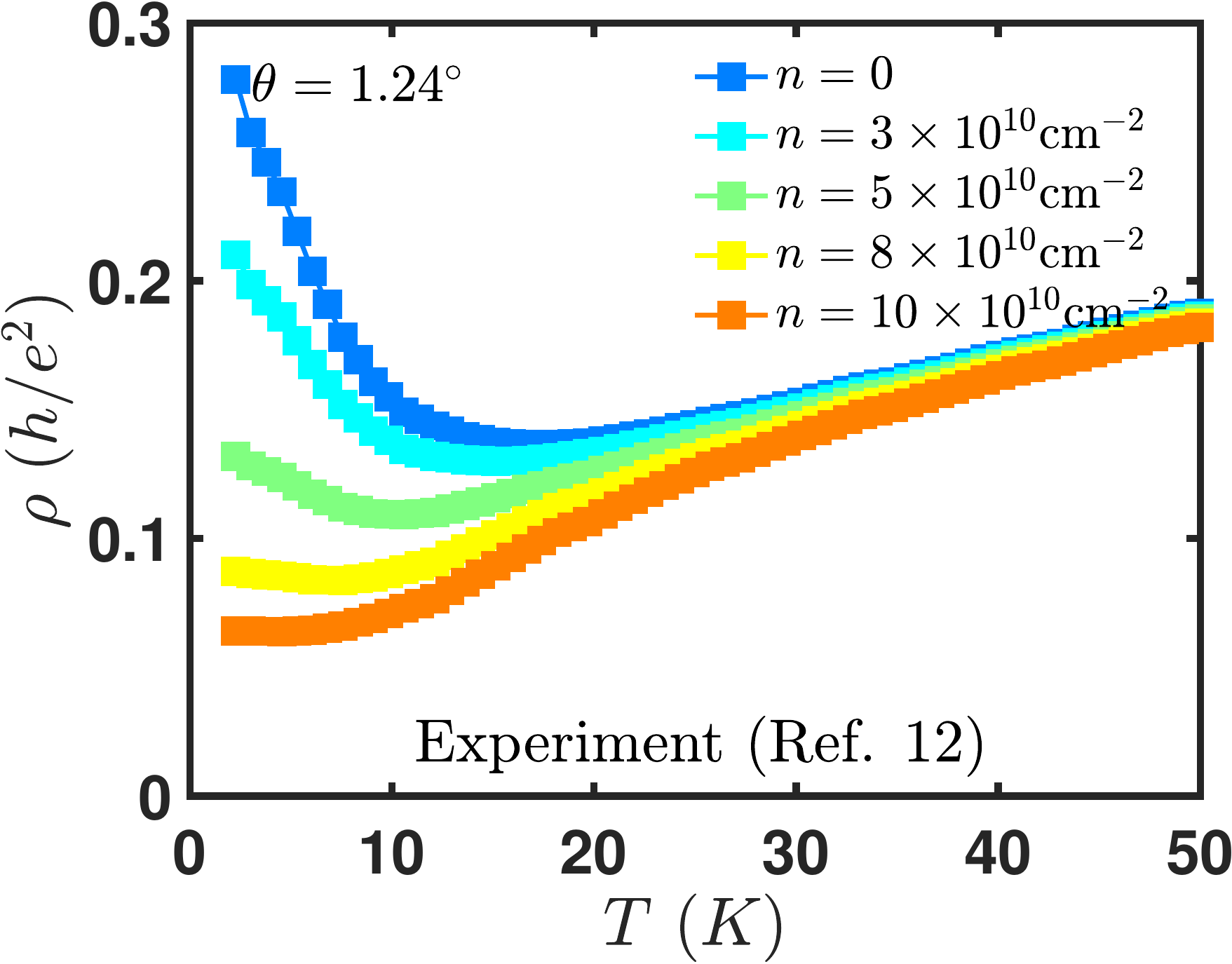}
\llap{\parbox[b]{6.4in}{\Large{(a)}\\\rule{0ex}{2.25in}}}
\includegraphics[scale=0.45]{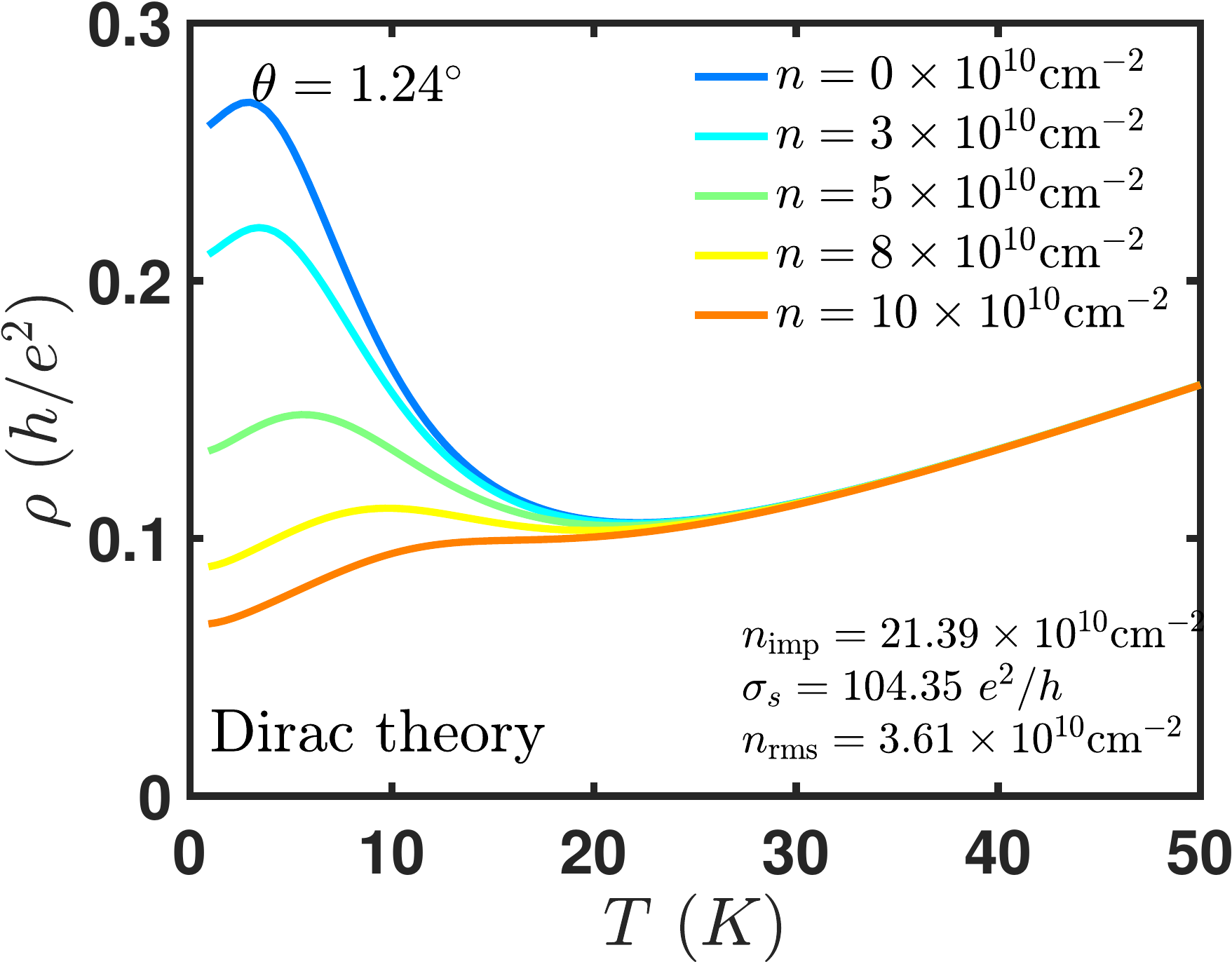}
\llap{\parbox[b]{6.4in}{\Large{(b)}\\\rule{0ex}{2.25in}}}
\caption{The non-monotonic temperature dependence at low temperatures and linear-in-$T$ dependence at intermediate temperatures seen in experiment is well captured within the Dirac theory of electron-(gauge) phonon scattering \cite{yudhistira2019gauge}. Resisitivity vs temperature from (a) Ref.~\cite{polshyn_2019_large} and (b) Dirac theory \cite{yudhistira2019gauge} at twist angle of $\theta=1.24^\circ$. The non-monotonicity at low temperatures is due to the crossover from impurity dominated transport to phonon dominated transport.\label{Fig:rho_exp_Dirac_vsT}}
\end{figure}

In Fig.~\ref{Fig:rho_exp_Dirac_vsn}, we show the comparison of experimental resistivity vs density from Ref.~\cite{polshyn_2019_large} to theoretical resistivity due to electron-phonon and electron-impurity scattering within Dirac model \cite{yudhistira2019gauge}. At low temperatures, electron-impurities dominates over electron-phonon interaction~\cite{yudhistira2019gauge}, therefore we only fit the experiment to the electron-impurity limited resistivity. These charged impurities also give rise to carrier density inhomogeneities~\cite{adam_self-consistent_2007}, which cures the otherwise divergent resistivity at the charge neutrality. We fixed the parameters from the effective medium theory (EMT)~\cite{rossi_effective_2009} fit to experimental resistivity at the lowest temperature, i.e.~$2$ K. In the fit, we have included short-range scattering component of the conductivity $\sigma_\mathrm{s}$~\cite{jang2008tuning}, charged impurity density $n_\mathrm{imp}$, and charge density fluctuations $n_\mathrm{rms}$. The first two parameters are used to calculate the Boltzmann-RPA conductivity $\sigma_\mathrm{B}[\sigma_\mathrm{s},n_\mathrm{imp}]$, which takes into account both the dominant scattering mechanism of screened Coulomb impurities and additional scattering mechanisms due to short-range scatterers, such as point defects and line defects. The parameter $n_\mathrm{rms}$ enters through the EMT equations. We find that the $1.24^\circ$ device of Ref.~\cite{polshyn_2019_large} has charged impurity density $n_\mathrm{imp}=2.1\times 10^{11}\mathrm{cm}^{-2}$, short ranged conductivity $\sigma_\mathrm{s}=104~e^2/h$ and charge density fluctuations $n_\mathrm{rms}=3.6\times 10^{10}\mathrm{cm}^{-2}$. At temperature of $10$ K and density of $\sim 10^{11}\mathrm{cm}^{-2}$, electron-phonon interaction start to produce noticeable influence on the resistivity. We observe the reversal of temperature dependence trend at $10$ K that is followed by linear-in-$T$ dependence at intermediate temperatures (see Fig.~\ref{Fig:rho_exp_Dirac_vsT}) and weaker density dependence at higher temperature in experiment. These were all predictions we made in our previous work~\cite{yudhistira2019gauge} that have been now shown experimentally to be correct.

In Figure.~\ref{fig1}, we compare the electron-phonon resistivity $\langle\rho\rangle=\sqrt{\rho_{xx}\rho_{yy}}$ from the two band effective model (left panel) with the experiment from Ref.~\cite{polshyn_2019_large} (middle panel) at twist angle of $\theta=1.1^\circ$. The fitting parameters are Fermi velocity $v_F$ and enhanced gauge field coupling constant $\tilde{\beta}_A$, which are obtained from fitting the temperature dependence of resistivity at density $n=10^{11}\mathrm{cm}^{-2}$ for fixed twist angle. The dependence of $v_F$ and $\tilde{\beta}_A$ on twist angle are plotted in Fig.~\ref{fig:parameter}a and \ref{fig:parameter}c, respectively. The same parameters are used to plot the temperature dependence of electron-phonon resistivity at higher density as well as its density dependence at several fixed temperatures (see the left panel of Fig.~\ref{fig1}b and \ref{fig1}a, respectively). We have fitted the electron-side and hole-side separately due to slight asymmetry between them. We find that in the low temperature regime, our results coincide with the Dirac model as expected, since the Hamiltonian in Eq.~\ref{CN} reduces to a Dirac Hamiltonian at low energies. However, in the high temperature regime only two band effective model agree with the experiment (see Fig.~\ref{fig1}). We see a clear saturation of the resistivity at higher temperatures as well as a weak density dependence in the electron and the hole side, as seen in experiment. Both of these features are not captured within the Dirac model. From fitting of the electron-phonon theory to the experiment (see Fig.~\ref{fig:parameter}c), we obtain $\beta_A$ around $2-6$ eV, which is in good agreement with the accepted values for the gauge field coupling constant in monolayer graphene \cite{sohier_phonon-limited_2014,kaasbjerg2012unraveling}. The fit values for the velocity ratio $v_F/v_0$ also agree well with theoretical estimates \cite{bistritzer2011moire}. Hence, we have developed a transport theory of tBG which explains all the salient features observed in the entire metallic regime (intermediate and high temperatures) of the experiment \cite{polshyn_2019_large}.

We also compared the Planckian resistivity to the same experiment in Fig.~\ref{fig1} (right panel and middle panel, respectively). The fitting parameters are Fermi velocity $v_F$ and Planckian strength $C$ (see Fig.~\ref{fig:parameter}b and \ref{fig:parameter}d, respectively). The parameters are obtained in similar manner with how we fit the electron-phonon theory. Similarly, they are used to plot the temperature dependence of Planckian resistivity at higher density as well as its dependence on density at several fixed temperatures (see the right panel of Fig.~\ref{fig1}b and \ref{fig1}a, respectively). Although it exhibit similar linear-in-$T$ behavior at low temperature and saturation at high temperature, its density dependence is much stronger than that of experiment, especially at low temperatures. The Fermi velocity $v_F$ extracted from the fit is somewhat larger than the theoretical prediction (see Fig.~\ref{fig:parameter}b). Moreover, we observe a violation of Planckian bound ($C>1$) for small twist angle. This rules out Planckian theory as the dominant mechanism of metallic transport in tBG.


\begin{figure*}[h!]
\begin{center}
\includegraphics[height=!,width=14cm]{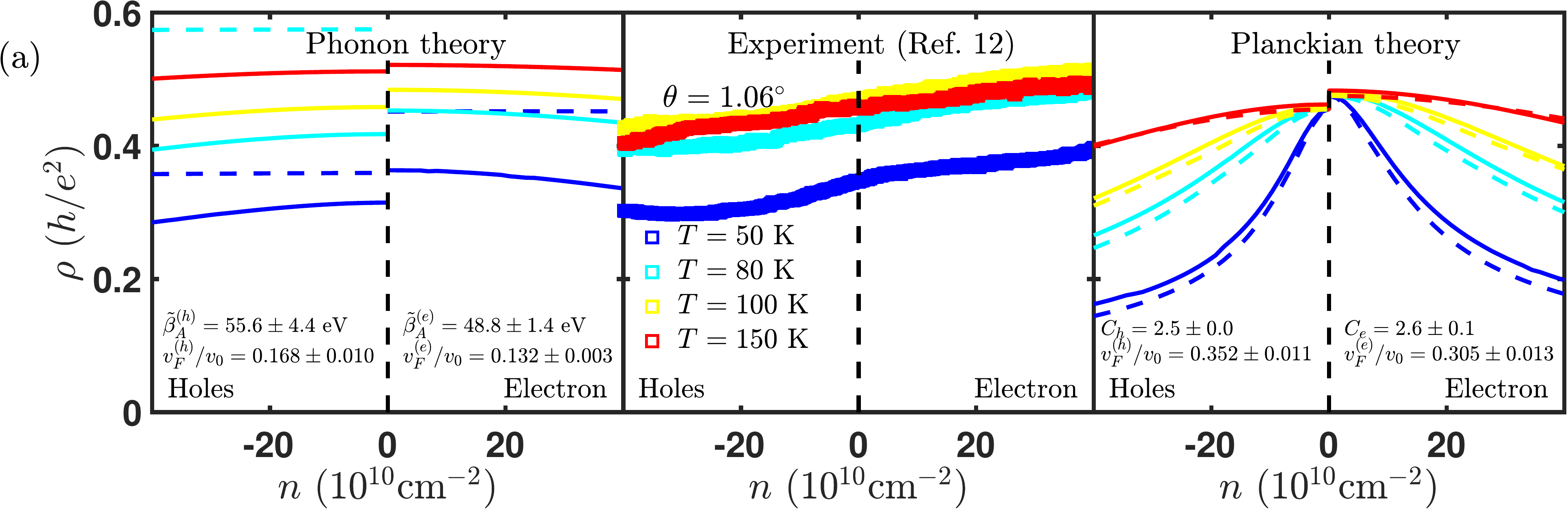}
\includegraphics[height=!,width=14cm]{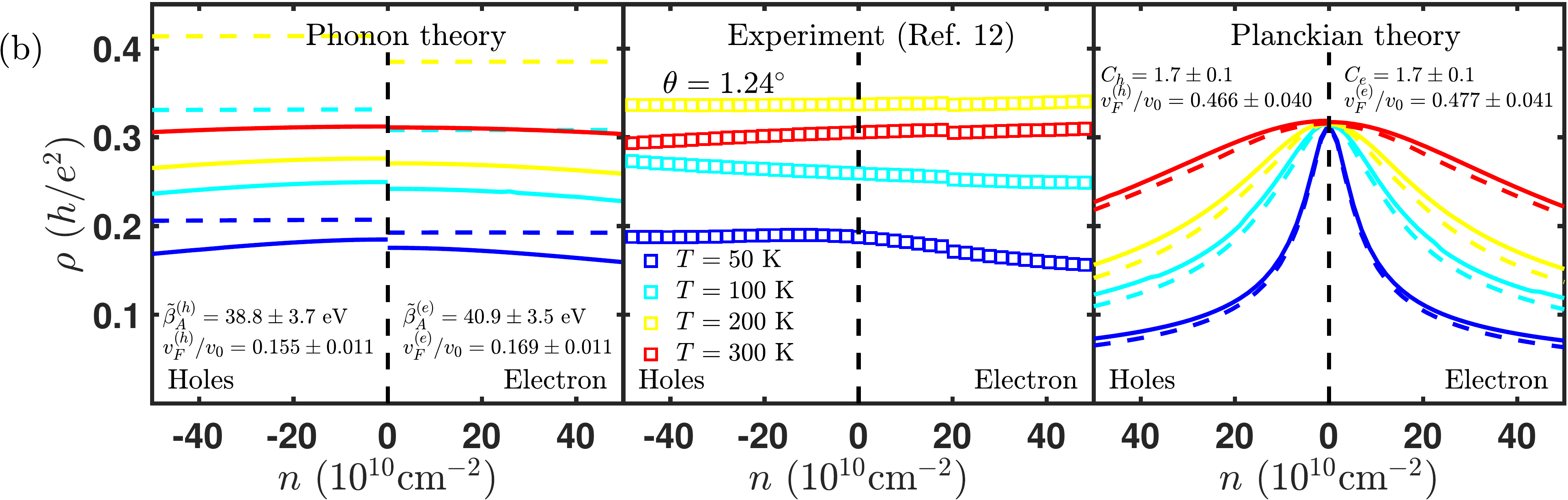}
\includegraphics[height=!,width=14cm]{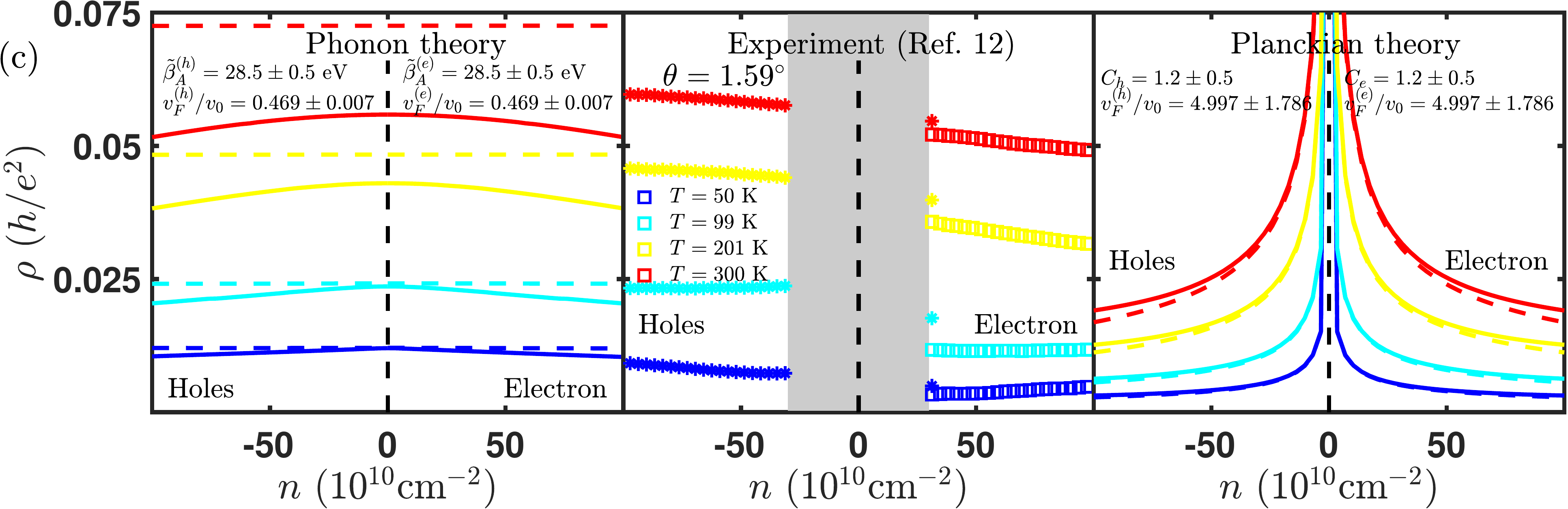}
\includegraphics[height=!,width=14cm]{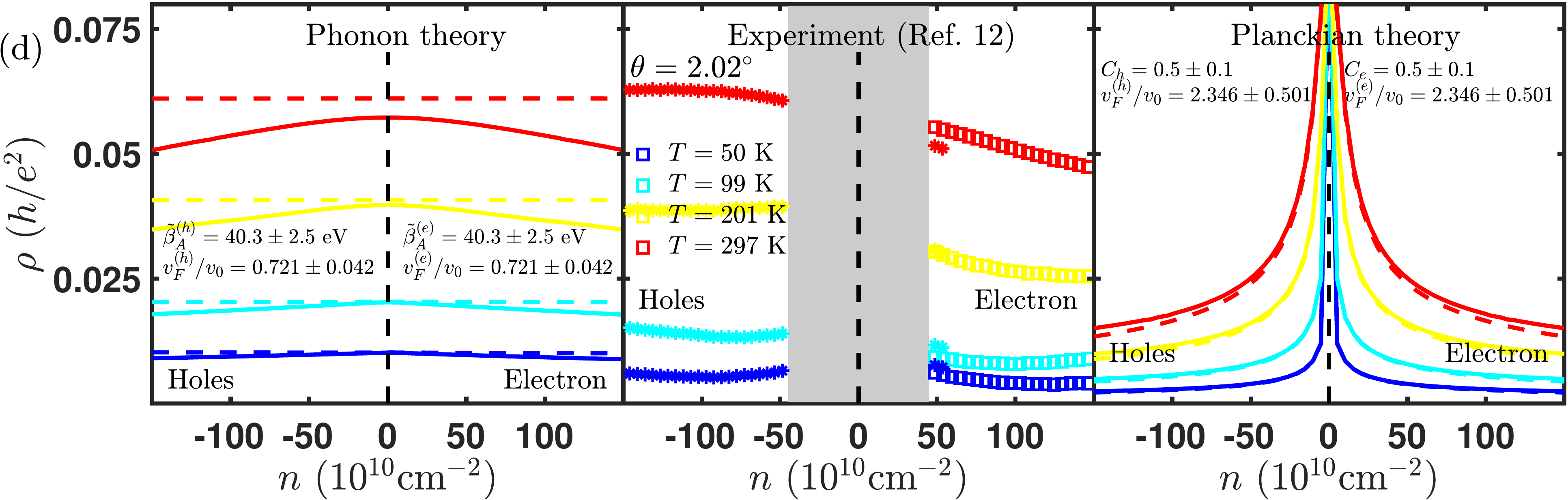}
\caption{(Color online) Resistivity vs charge density of experiment in Ref.~\cite{polshyn_2019_large} (middle panel) compared to electron-phonon theory (left panel) and Planckian theory (right panel) at twist angle of (a) $1.06^\circ$ (b) $1.24^\circ$ (c) $1.59^\circ$ and (d) $2.02^\circ$.}
\end{center}
\end{figure*}

\begin{figure*}
\begin{center}
\includegraphics[height=!,width=14cm]{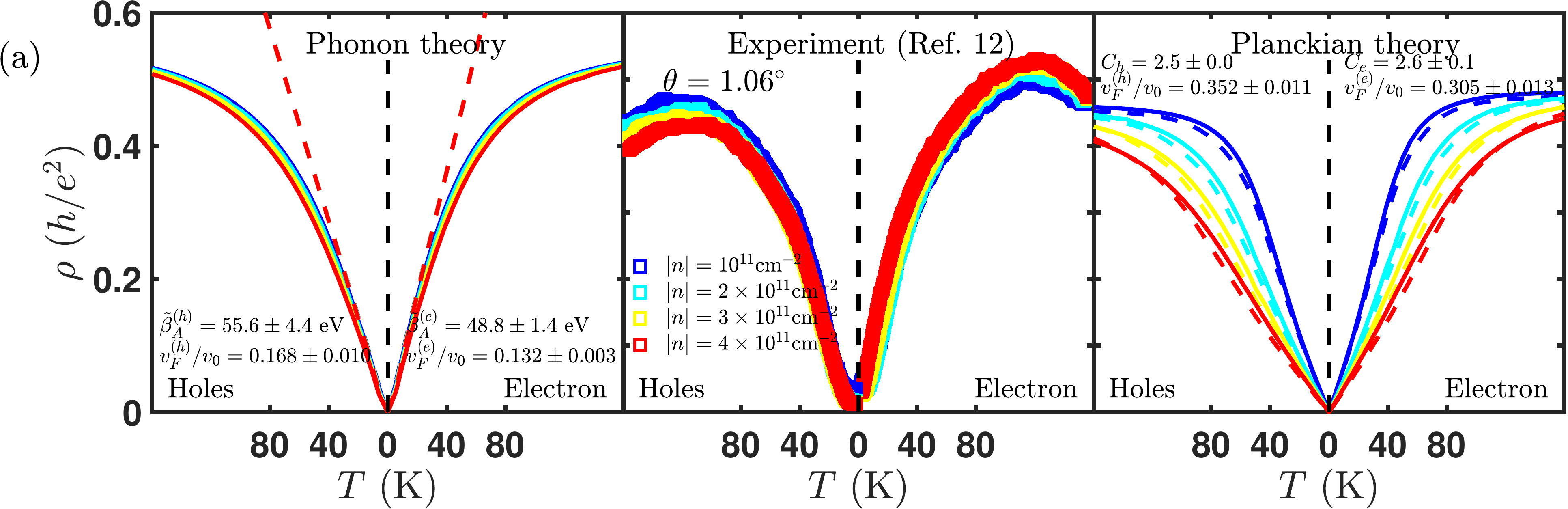}
\includegraphics[height=!,width=14cm]{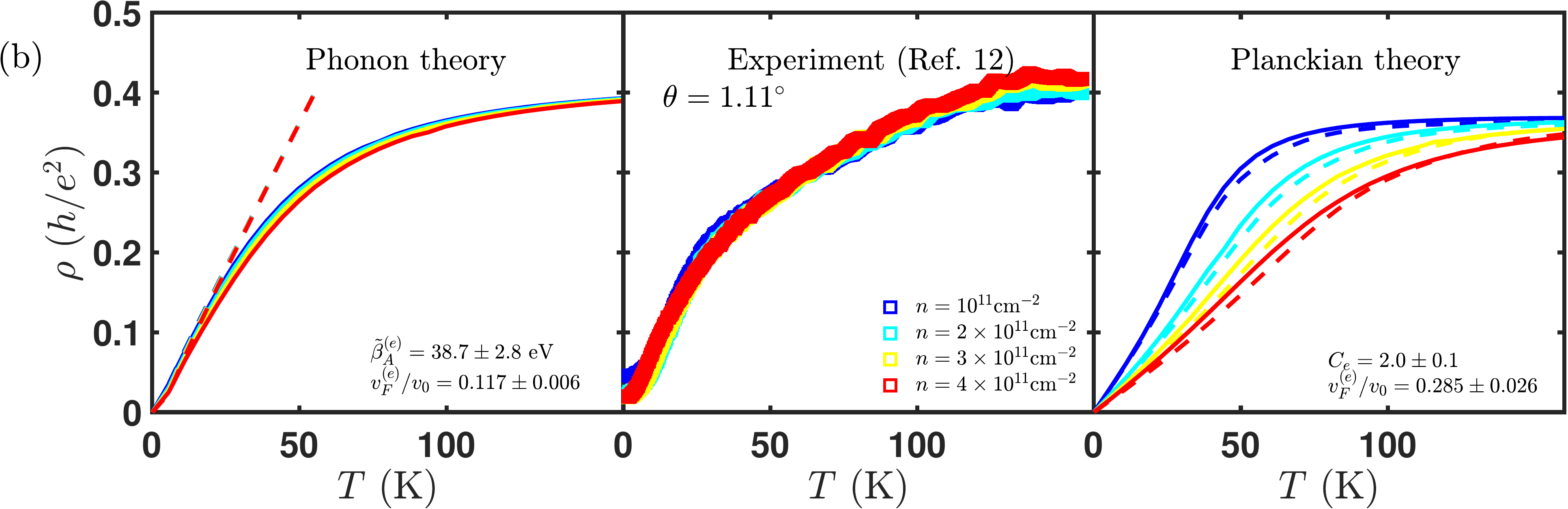}
\includegraphics[height=!,width=14cm]{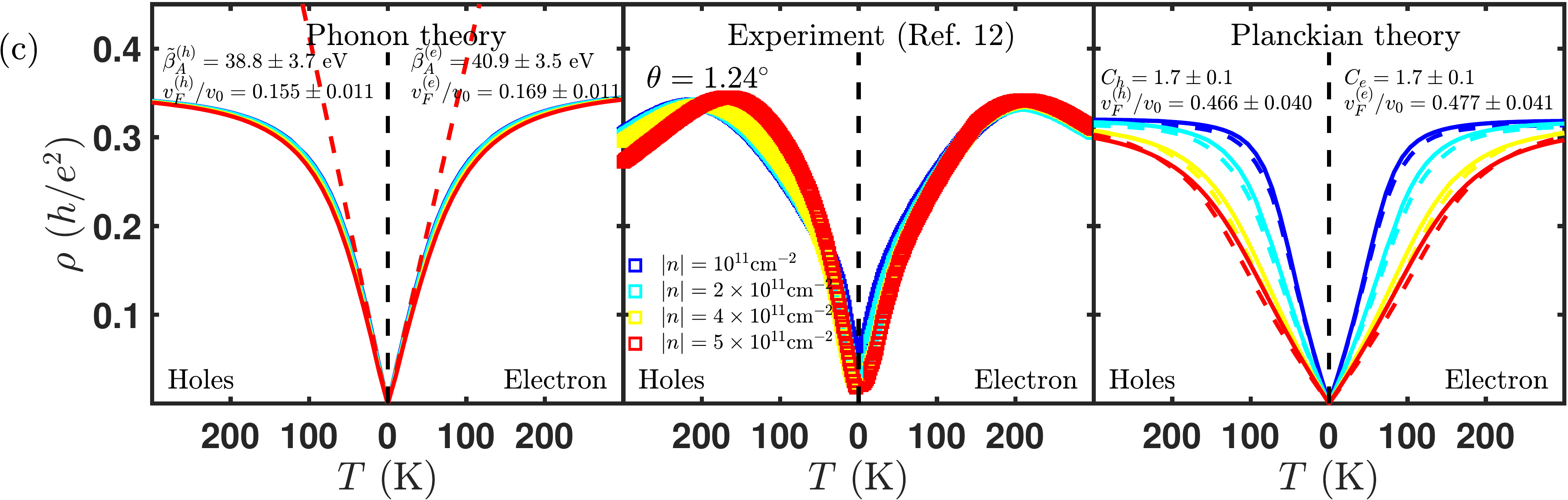}
\includegraphics[height=!,width=14cm]{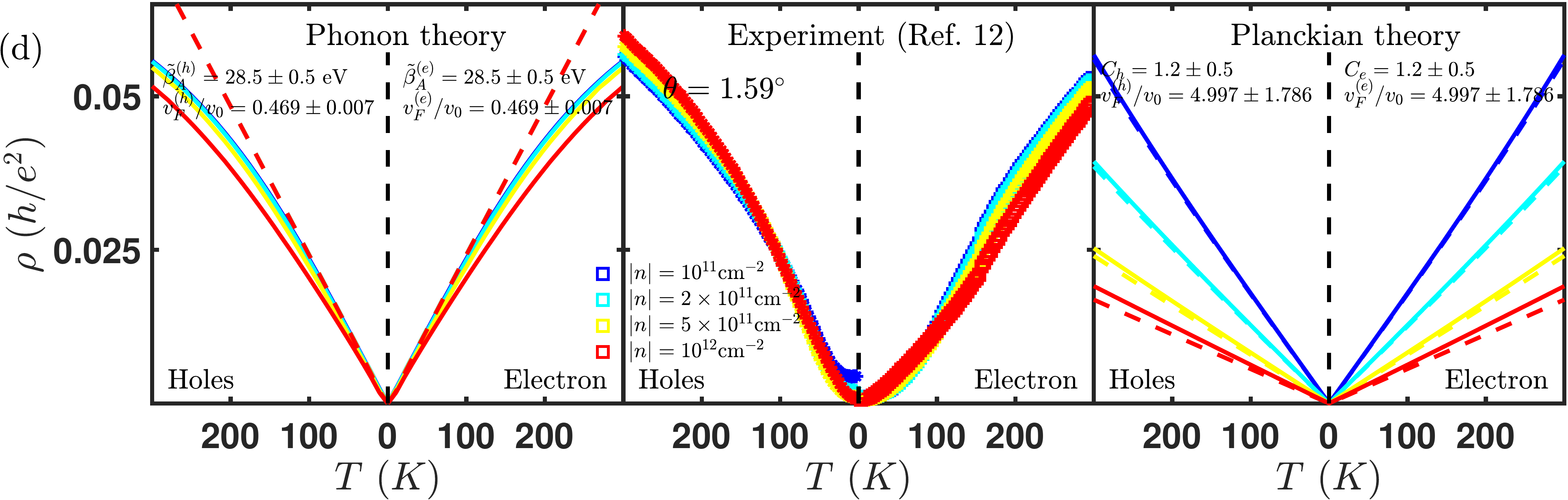}
\includegraphics[height=!,width=14cm]{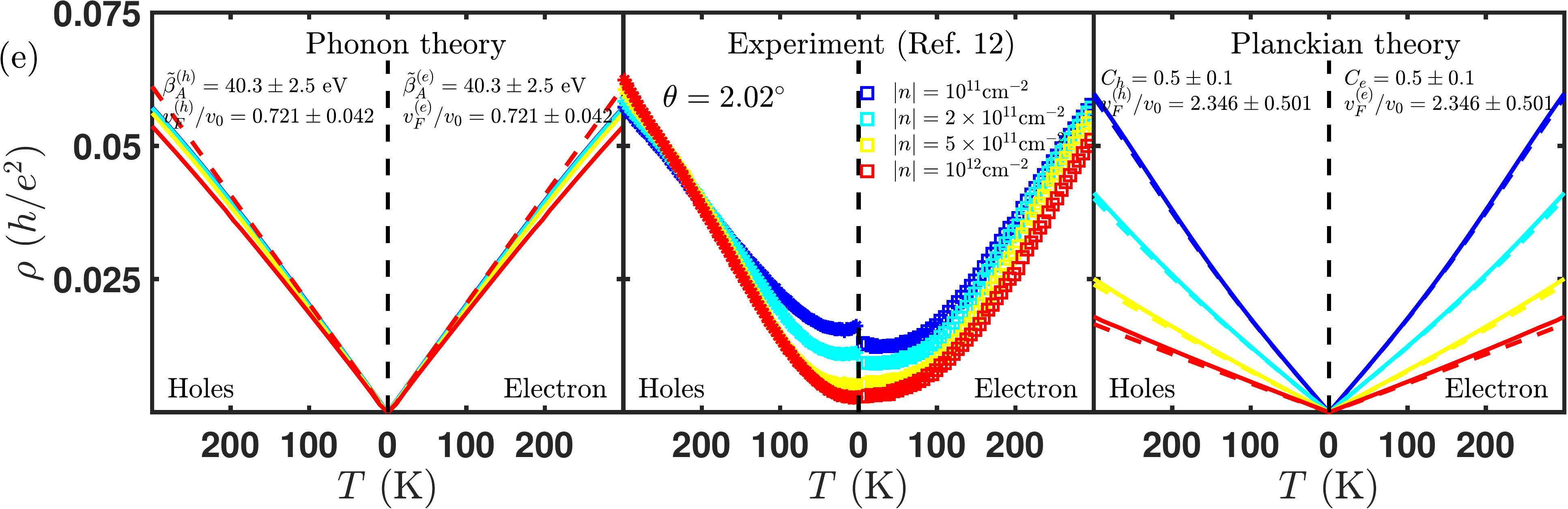}
\caption{(Color online) Resistivity vs temperature of experiment in Ref.~\cite{polshyn_2019_large} (middle panel) compared to electron-phonon theory (left panel) and Planckian theory (right panel) at twist angle of (a) $1.06^\circ$ (b) $1.11^\circ$ (c) $1.24^\circ$ (d) $1.59^\circ$ and (e) $2.02^\circ$.}
\end{center}
\end{figure*}

\end{widetext}

\end{document}